\documentclass[a4paper]{article}

\usepackage[utf8]{inputenc}
\usepackage{xspace}
\usepackage{algorithm}
\usepackage[noend]{algpseudocode}
\usepackage{amsmath}
\usepackage{amsthm}
\usepackage{amssymb}
\usepackage{textcomp}
\usepackage{stmaryrd}
\usepackage{varwidth}
\usepackage{graphicx}
\usepackage{hyperref}
\usepackage{fullpage}

\usepackage{wrapfig}
\usepackage{bbold}
\usepackage{pgfplots}
\usepackage{xspace}

\usepackage{calc}
\usepackage{paralist}
\usepackage{enumitem}
\usepackage{balance}
\usepackage{footnote}

\usepackage{dsfont}

\usepackage{soul}

\usepackage{tikz}
\usetikzlibrary{automata, graphs,positioning,chains,arrows,snakes,decorations.pathmorphing,calc}

\usepackage{fancyvrb}

\theoremstyle{definition}
\newtheorem{example}{Example}[section]
\newtheorem{theorem}{Theorem}[section]
\newtheorem{lemma}{Lemma}[section]

\usepackage
{todonotes}

\newcommand{\cer}{CER\xspace}
\newcommand{\cel}{CEL\xspace}
\newcommand{\acel}{ACEL\xspace}
\newcommand{\TRUE}{\mathtt{TRUE}}

\newcommand{\dset}{\mathbf{D}}
\newcommand{\aset}{\mathbf{A}}
\newcommand{\asetSchema}{\mathbf{A}_{\Sigma}}
\newcommand{\asetNotSchema}{\bar{\mathbf{A}}_{\Sigma}}
\newcommand{\tset}{\mathbf{T}}

\newcommand{\eset}{\mathbf{E}}

\newcommand{\type}{\operatorname{type}}
\newcommand{\nullvalue}{\texttt{NULL}}
\newcommand{\Stream}{\mathcal{S}}
\newcommand{\Streamstocks}{\mathcal{S}_{\operatorname{Stocks}}}

\newcommand{\Schemastocks}{\Sigma_{\operatorname{Stocks}}}
\newcommand{\schema}{\texttt{schema}}
\newcommand{\ISO}{\text{ISO}}

\newcommand{\ctime}{\operatorname{time}}
\newcommand{\cstart}{\operatorname{start}}
\newcommand{\cend}{\operatorname{end}}

\newcommand{\tuples}{\operatorname{Tuples}}

\newcommand{\assigns}{\operatorname{Asg}}

\newcommand{\dom}{\operatorname{dom}}
\newcommand{\img}{\operatorname{img}}
\newcommand{\pow}[1]{\mathcal{P}({#1})}
\newcommand{\bagpow}[1]{\mathcal{P}_{bags}({#1})}

\newcommand{\ba}{\textbf{a}}
\newcommand{\bb}{\textbf{b}}
\newcommand{\bc}{\textbf{c}}
\newcommand{\bd}{\textbf{d}}
\newcommand{\bn}{\textbf{n}}
\newcommand{\bm}{\textbf{m}}
\newcommand{\bp}{\textbf{p}}

\newcommand{\valvar}{\textbf{value}}

\newcommand{\zero}{\mathds{O}}
\newcommand{\one}{\mathds{1}}
\newcommand{\expr}{\operatorname{Expr}}

\newcommand{\cA}{\mathcal{A}}

\newcommand{\cD}{\mathcal{D}}

\newcommand{\cS}{\mathcal{S}}

\newcommand{\bbN}{\mathbb{N}}

\newcommand{\pset}{\mathbf{P}}

\newcommand{\xset}{\mathbf{X}}
\newcommand{\sem}[1]{{\llbracket{}{#1}\rrbracket}}

\newcommand{\bleft}{\{\!\!\{}
\newcommand{\bright}{\}\!\!\}}

\newcommand{\kAS}{~\operatorname{ AS }~}
\newcommand{\kFILTER}{~\operatorname{ FILTER }~}
\newcommand{\kOR}{~\operatorname{ OR }~}
\newcommand{\kAND}{~\operatorname{ AND }~}
\newcommand{\kSEQ}{\,;\,}
\newcommand{\kSSEQ}{\,:\,}
\newcommand{\kITER}{+}
\newcommand{\kSITER}{\oplus}
\newcommand{\kPROJ}{\pi}

\newcommand{\auxsem}[1]{{\llceil{#1}\rrfloor}}
\newcommand{\semaux}[1]{\sem{#1}}

\newcommand{\sq}{\,;\,}
\newcommand{\ssq}{:}

\newcommand{\fsq}{\,:\,}

\newcommand{\agg}{\mathtt{Agg}}
\newcommand{\nxt}{\mathtt{NEXT}}

\newcommand{\sumtext}{\operatorname{sum}}
\renewcommand{\min}{\operatorname{min}}
\renewcommand{\max}{\operatorname{max}}
\newcommand{\cnt}{\operatorname{count}}
\newcommand{\avg}{\operatorname{avg}}
\newcommand{\rng}{\operatorname{range}}

\newcommand{\CEA}{CEA\xspace}
\newcommand{\ACEA}{ACEA\xspace}

\newcommand{\trans}[2][]{\raisebox{-1pt}[10pt][0pt]{$\overset{#2}{\underset{^{#1}}{\raisebox{0pt}[3pt][0pt]{$\relbar\mspace{-8mu}\longrightarrow$}}}$}}

\newcommand{\llltrans}[2][]{\raisebox{-1pt}[10pt][0pt]{$\overset{#2}{\underset{^{#1}}{\raisebox{0pt}[3pt][0pt]{$\relbar\mspace{-8mu}\relbar\joinrel\relbar\joinrel\relbar\joinrel\relbar\joinrel\longrightarrow$}}}$}}
\newcommand{\lltrans}[2][]{\raisebox{-1pt}[10pt][0pt]{$\overset{#2}{\underset{^{#1}}{\raisebox{0pt}[3pt][0pt]{$\relbar\mspace{-8mu}\relbar\joinrel\relbar\joinrel\longrightarrow$}}}$}}

\newcommand{\att}{\operatorname{Att}}
\newcommand{\attin}{\operatorname{Att}_{\operatorname{in}}}
\newcommand{\attout}{\operatorname{Att}_{\operatorname{out}}}

\newcommand{\TR}{\operatorname{Ren}}

\renewcommand{\paragraph}[1]{\smallskip
	\noindent\textbf{#1}.}

\algrenewcommand\algorithmicindent{1.3em}%
\algnewcommand\algorithmicforeach{\textbf{for each}}
\algdef{S}[FOR]{ForEach}[1]{\algorithmicforeach\ #1\ \algorithmicdo}

\newcommand{\ssd}[3]{\arraycolsep=0pt
	\left[
	\renewcommand{\arraystretch}{1}
	\begin{array}{c}
          \textnormal{\textsf{#1}} \\ \textnormal{\textsf{#2}} \\ #3
	\end{array}
	\right]}

\title{A formal query language and automata model \\
	for aggregation in complex event recognition} 

\author{Pierre Bourhis$^1$ \and Cristian Riveros$^2$ \and Amaranta Salas$^2$}

\date{
	$^1$University of Lille, CNRS \\ \texttt{pierre.bourhis@univ-lille.fr}\\%
	$^2$Pontificia Universidad Católica de Chile \\ \texttt{\{cristian.riveros, afsalas\}@uc.cl} \\[2ex]%
}

\begin{document}
	
	\maketitle

\begin{abstract}

Complex Event Recognition (CER) systems are used to identify complex patterns in event streams, such as those found in stock markets, sensor networks, and other similar applications. An important task in such patterns is aggregation, which involves summarizing a set of values into a single value using an algebraic function, such as the maximum, sum, or average, among others. Despite the relevance of this task, query languages in CER typically support aggregation in a restricted syntactic form, and their semantics are generally undefined. 

In this work, we present a first step toward formalizing a query language with aggregation for CER. We propose to extend Complex Event Logic (CEL), a formal query language for CER, with aggregation operations. This task requires revisiting the semantics of CEL, using a new semantics based on bags of tuples instead of sets of positions. Then, we present an extension of CEL, called Aggregation CEL (ACEL), which introduces an aggregation operator for any commutative monoid operation. The operator can be freely composed with previous CEL operators, allowing users to define complex queries and patterns. We showcase several queries in practice where ACEL proves to be natural for specifying them. From the computational side, we present a novel automata model, called Aggregation Complex Event Automata (ACEA), that extends the previous proposal of Complex Event Automata (CEA) with aggregation and filtering features. Moreover, we demonstrate that every query in ACEL can be expressed in ACEA, illustrating the effectiveness of our computational model. Finally, we study the expressiveness of ACEA through the lens of ACEL, showing that the automata model is more expressive than ACEL. 

\end{abstract} 	
	\section{Introduction}\label{sec:introduction}

Complex Event Recognition (\cer) systems are a group of data stream management systems for the detection of special events in real-time, called complex events, that satisfy a pattern, considering their position, the order, and other constraints between them \cite{giatrakos2020complex, cugola2012processing}. Some examples of its use are maritime monitoring \cite{pitsikalis2019composite}, network intrusion detection \cite{mukherjee1994network}, industrial control systems \cite{groover2016automation}, and real-time analytics \cite{sahay2008real}. In the literature, people have proposed multiple systems and query languages based on different formalisms for approaching complex events, such as automata-based, logic-based, tree-based, or a combination of them \cite{giatrakos2020complex}. Examples of \cer systems developed in academic and industrial contexts include SASE \cite{wu2006high}, EsperTech \cite{espertech}, and CORE~\cite{BucchiGQRV22,BossonneyBCLRV25}, among others.

A problem in CER systems is that their query languages, used to declare complex events, are, unfortunately, underspecified with respect to both their syntax and semantics. 
As observed in previous works~\cite{zimmer1999semantics,galton2002two,CugolaM10,ArtikisMUVW17}, CER query languages in systems typically lack a simple, compositional, and denotational semantics. In general, its semantics is defined indirectly through examples \cite{adi2004amit,cugola2009raced,luckham1996rapide}, or by translation into evaluation models \cite{pietzuch2003framework,schultz2009distributed,white2007next}. Recently, this issue in CER systems has been studied more thoroughly, and a query language that has successfully defined the semantics of several \cer operators is \emph{Complex Event Logic} (\cel) \cite{GrezRU19,GrezRUV21}, alongside a computational model called \emph{Complex Event Automata} (\CEA), which is based on the theory of finite state transducers and symbolic automata.

An open problem in formalizing \cer query language is that several interesting queries in practice include \emph{aggregation}, which previous proposals have not addressed. Aggregation refers to any subprocess in a query that combines and merges several (most often numerical) values into a single one \cite{grabisch2009aggregation}, such as taking the average, the sum, or the maximum of a list of values. Examples of CER systems that used aggregation are SASE \cite{DiaoIG07, SASEcomplexity}, EsperTech \cite{espertech}, GLORIA \cite{Ma2022}, GRETA \cite{Poppe2020}, and others~\cite{giatrakos2020complex, cugola2012processing}. 
For illustrating a prototypical query with aggregation, consider the following (simplified) query from SASE ~\cite[p.3]{SASEcomplexity}:
\begin{verbatim}
	Q1: PATTERN seq(JobStart a, Mapper+ b[ ], JobEnd c)
	WHERE a.job_id = b[i].job_id and a.job_id = c.job_id
	RETURN AVG(b[ ].period), MAX(b[ ].period)
\end{verbatim}    
Intuitively, the previous query aims to retrieve the average and maximum periods from a list of running times of mappers. For this, it looks for events from the stream, in the `\texttt{PATTERN}' clause that match the pattern: a JobStart typed event, followed by one or more Mapper events, and finally a JobEnd event. In turn, it uses the `\texttt{WHERE}' clause to ensure that each event matching the pattern has the same id attribute, and finally, it returns the average and maximum of the events that meet the conditions. 

Previous proposals for formalizing CER query languages do not include such queries with aggregation, as they must not only detect and retrieve complex events but also produce new events and values. In particular, aggregation queries cannot be defined by logics like \cel or computational models like \CEA, or any other formalization of \cer as it is currently defined. These issues imply that CER query languages with aggregation are difficult to compare, unclear how to compose queries, and difficult to evaluate (i.e., without knowing the real meaning of a query). Furthermore, computational models for compiling queries with aggregation are not well understood, and systems usually rely on ad-hoc evaluation strategies suitable for specific queries and patterns. 

In this work, we propose an extension of the logic \cel, and its corresponding computational model \CEA, to express queries with aggregation, which we call \emph{Aggregation Complex Event Logic} (\acel) and \emph{Aggregation Complex Event Automata} (\ACEA), respectively. Our main goal is to design a logic and computational model, with a formal semantics that formalizes aggregation in CER and serves as a base for all \cer languages.

For extending \cel with aggregation, we need to revisit its semantics. One of the first problems to arise with the current semantics of CEL is that a CEL formula retrieves the positions in the streams that fire the complex events, but it does not allow the creation of new values or events. For this reason, we propose a new, equivalent semantics for CEL that returns events instead of positions. Additionally, since we also need to maintain duplicates for aggregation, we extend the semantics by using bags of events instead of sets. We then prove that the new semantics is equivalent to the previous one. Interestingly, the new semantics enable us to define new relevant operators for CER, such as attribute projection.%

To formalize the aggregation in \cer,  we consider a general setting of aggregation based on \emph{aggregate functions} \cite{jesus2014survey,grabisch2009aggregation}; these are functions that go from a bag of values to single values, and they aim to summarize information (like count, sum, etc). By using this general framework of aggregation functions, they support our proposal in providing a general framework for aggregation in CER. Furthermore, we introduce an operator $\agg_{Y (\bb \gets \otimes X(\ba))}$ for variables named $X$ and $Y$, attributes named $\ba$ and $\bb$, and aggregation function $\otimes$, which takes a bag of events stored in $X$ and $\otimes$-operates it corresponding attribute $\ba$, storing the result in another attribute $\bb$ of an event in another variable $Y$. We formally define its syntax and semantics in Section $\ref{sec:aggregation}$. An advantage of this definition is that we can compose the $\agg$ operator and every other operator in \cel. We show that most CER queries with aggregation from previous works are definable with \acel.

An advantage of \cel is that one can characterize its expressive power with the so-called Complex Event Automata (\CEA); specifically, that for every \cel formula, there exists an equivalent \CEA, and vice versa. The practical relevance of this result is that \cel is useful for users to define queries, where \CEA is useful for systems to evaluate them. 
In this work, we aim to achieve an equivalent result, so our next step was to find a machine model that can extend \CEA and formally define \acel. We introduce an automata model with aggregation for \acel, which we call \emph{Aggregation Complex Event Automata} (\ACEA), an extension of \CEA with registers to aggregate values. This extension employs the same concept of operating values in transitions and maintaining registers as cost register automata \cite{alur2013cra}. Specifically, in each transition, an \ACEA takes an event and updates its register with those new values. Then, it performs an operation based on the assignments, checks if it satisfies a predicate, and finally, it creates a new tuple with the aggregated values. One of our main results is that we can compile every \acel formula into an \ACEA, namely, we can prove that the expressive power of \acel is a subset of \ACEA.

\paragraph{Outline}
We present the preliminaries in Section \ref{sec:preliminaries}. In Section \ref{sec:cel}, we discuss the necessary changes in \cel semantics and we show a new operator, projection by attribute. 
In Section~\ref{sec:aggregation-func}, we discuss the setting of aggregate functions. In Section \ref{sec:aggregation}, we formally introduce \acel, 
and we introduce \ACEA in Section \ref{sec:acea}, to study the compilation of \cel formulas into \ACEA and its equivalence with \CEA. We conclude and discuss future work in Section \ref{sec:conclusions}.

	\section{Preliminaries}\label{sec:preliminaries}

\paragraph{Sets, intervals, and mappings}  Given a set $A$, we denote by $\pow{A}$ the set of all finite subsets of $A$. We denote by $\bbN$ the natural numbers. Given $n,m \in \bbN$ with $n \leq m$, we denote by $[n]$ the set $\{1, \ldots, n\}$ and by $[n..m]$ the interval $\{n, n+1, \ldots, m\}$ over $\bbN$. As usual, we write $f: A \rightarrow B$ to denote a \emph{function} $f$ from the set $A$ to $B$ where every element in $A$ has an image. A \emph{mapping} $M$ is a partial function that maps a finite number of elements from $A$ to elements over $B$. We write $M: A \mapsto B$ to denote a mapping $M$ from $A$ to~$B$. We denote by $\dom(M)$ the domain of $M$ (i.e., all $a \in A$ such that $M(a)$ is defined), and by $\img(M)$ the image of $M$. 
We will usually use the notation $[a_1 \mapsto b_1, \ldots, a_k \mapsto b_k]$ to define a mapping $M$ with $\dom(M) = \{a_1, \ldots, a_k\}$, $\img(M) = \{b_1, \ldots, b_k\}$ and $M(a_i) = b_i$ for every $i \in [k]$. Furthermore, for a map $M$ and $a \notin \dom(M)$ we write $[M, a \mapsto b]$ to specify a new map $M'$ that extends $M$ mapping $a$ to $b$.%

\paragraph{Bags} A \emph{bag} or \emph{multiset} (with own identity) $B$ is a mapping $B: I \mapsto U$ where $I = \dom(B)$ is a finite set of identifiers (or ids) and $U = \img(B)$ is the underliying set of the bag. Given any bag $B$, we refer to these components as $I(B)$ and $U(B)$, respectively. For example, a bag $B = \bleft a, a, b \bright$ (where $a$ is repeated twice) can be represented with a mapping $B_0 =[1 \mapsto a, 2 \mapsto a, 3 \mapsto b]$ where $I(B_0) = \{1, 2, 3\}$ and $U(B_0) = \{a,b\}$. In general, we will use the standard notation for bags $\bleft a_1, \ldots, a_{n} \bright$ to denote the bag $B$ whose identifiers are $I(B) = \{1, \ldots, n\}$ and $B(i) = a_i$ for each $i \in I(B)$. We will use $\uplus$ to refer to the union of bags.
Further, we define $\bagpow{A}$ as the set of all finite bags that one can form from a set $A$.

\paragraph{Computational model}
We assume the model of \emph{random access machines (RAM)} with uniform cost measure, and addition and subtraction as basic operations \cite{Aho1974design}. This implies, for example, that the access to a lookup table (i.e., a table indexed by a key) takes constant time. These are common assumptions in the literature of the area \cite{BucchiGQRV22,Segoufin2013enumerating}.
 	
	\section{Revisiting the semantics of Complex Event Logic}\label{sec:cel}

In this section, we revisit the semantics of CEL~\cite{GrezRUV21} and present a new semantics based on tuples instead of positions. We then prove the equivalence between the two versions. This new semantics allows, for example, the definition of a new operator for CEL, called \emph{attribute-projection}, which cannot be defined with the old semantics. Furthermore, the new semantics is crucial to introduce aggregation in CEL in the next section.

\paragraph{Events and streams} We fix a countably infinite set of \emph{attribute names} $\aset$ and a countably infinite of \emph{data values}  $\dset$  (e.g. integers, strings). 
An (untyped) \emph{event} $e$ is a pair $(M, i)$ such that $M: \aset \mapsto \dset$ maps attribute names from~$\aset$ to data values in~$\dset$, and $i \in \bbN$ is the time of the event\cite{GarciaR25} (we prefer to use \emph{discrete time}, which is enough for our purposes).
Intuitively, $M$ defines the data of the event (i.e., as a tuple).  
We denote by $e(\ba) \in \dset$ the value of the attribute $\ba \in \aset$ assigned by $M$ (i.e., $e(\ba) = M(\ba)$).
If $e$ is not defined on attribute $\ba$, then we write $e(\ba) = \nullvalue$. 
Furthermore, for the sake of simplification we also denote $e(\ctime) = i$ (note, however, that $\ctime$ is not an attribute).
We define by $\att(e)$ the set of attributes of $e$, namely, $\att(e) =  \dom(M)$.
We write $\eset$ to denote the \emph{set of all events} over attributes names $\aset$ and data values $\dset$.
We will usually use bold letters $\ba$, $\bb$, and $\bc$ to denote attribute names in $\aset$ and (normal) letters $a$, $b$, and $c$ to denote data values in $\dset$.

Fix now a finite set of \emph{event types} $\tset$ and assume that $\tset \subseteq \dset$ and $\nullvalue \in \dset$. 
In this work, we assume the existence of a distinguished attribute $\type$ that defines the \emph{type} of an event. Specifically, let $\type$ be an attribute such that $\type \in \aset$.
For every event $e$, we assume that $\type \in \att(e)$ and $e(\type) \in \tset \cup \{\nullvalue\}$ is the type of $e$. Notice that $e$ could be \emph{typed} (i.e., $e(\type) \in \tset$) or \emph{untyped} in which case we have $e(\type) = \nullvalue$. 
A \emph{schema} $\Sigma$ is a function $\Sigma: \tset \rightarrow \pow{\asetSchema}$ where $\asetSchema \subseteq \aset$ is a finite set of attributes. 
We say that an event $e$ satisfies the schema $\Sigma$ if, and only if, $e$ is a typed event and $\att(e) = \Sigma(e(\type)) \cup \{\type\}$. In particular, untyped events do not satisfy a schema by definition.

Let $\Sigma: \tset \rightarrow \pow{\asetSchema}$ be a schema. A \emph{stream} over a schema $\Sigma$ is an (arbitrary long) sequence $\Stream = e_1 e_2 \ldots e_{n}$ of typed events such that, for every $i \in [n]$, it holds that $e$ satisfies~$\Sigma$ and $e_i(\ctime) = i$. 
In other words, a stream consists of typed events according to $\Sigma$ and every time of an event is the position in the stream.
Note that we defined the type and time of an event for its later use in the semantics of CEL. The first will allow us to know the attributes of each event when we compile CEL into an automata model, and the second will allow us to differentiate between tuples by adding its \emph{origin} in the stream \cite{bojanczyk2014transducers}.%

\begin{example} \label{ex:stream}
	As a running example, consider that we have a stream $\Streamstocks$ that is emitting buy and sell events of particular stocks~\cite{BucchiGQRV22}. Here, we assume a schema $\Schemastocks$ with attributes $\mathsf{name}$ and $\mathsf{price}$ that represents the name of the stock (e.g., INTL for intel) and its price (e.g., US\$80), respectively. We have two types, called \texttt{BUY} and \texttt{SELL}, and $\Schemastocks(\texttt{BUY}) = \Schemastocks(\texttt{SELL}) = \{\mathsf{name}, \mathsf{price}\}$. A possible stream $\Streamstocks$ could be the following:
	\begin{center}
		\small
		\begin{tikzpicture}[->,>=stealth, semithick, auto, initial text= {}, initial distance= {3mm}, accepting distance= {4mm}, node distance=1.9cm, semithick]
			\tikzstyle{every state}=[draw=black,text=black,inner sep=0pt, minimum size=5mm]

			\begin{scope}[yshift=-2.2cm, node distance=1.1cm]
				\node[label=0] (1)             {$\scriptsize \ssd{SELL}{MSFT}{101}$};
				\node[label=1,right of=1] (2) {$\scriptsize \ssd{SELL}{MSFT}{102}$};
				\node[label=2,right of=2] (3) {$\scriptsize \ssd{SELL}{INTL}{80}$};
				\node[label=3,right of=3] (4) {$\scriptsize \ssd{BUY}{INTL}{80}$};
				\node[label=4,right of=4] (5) {$\scriptsize \ssd{SELL}{AMZN}{1900}$};
				\node[label=5,right of=5] (6) {$\scriptsize \ssd{SELL}{INTL}{81}$};
				\node[label=6,right of=6] (7) {$\scriptsize \ssd{BUY}{AMZN}{1920}$};
				\node[label=7,right of=7] (8) {$\scriptsize \ssd{BUY}{MSFT}{101}$};
				\node[label=8,right of=8] (9) {$\scriptsize \ssd{BUY}{INTL}{79}$};
				\node[label=9,right of=9] (10) {$\scriptsize \ssd{SELL}{INTL}{80}$};
				\node at ($(1) + (-1.35, 0)$) {$\Streamstocks$:};
			\end{scope}
			
		\end{tikzpicture}
	\end{center}
	Note that each event contains a type (i.e., \texttt{BUY} or \texttt{SELL}), its attributes values (i.e., $\mathsf{name}$ and $\mathsf{price}$) and its time (i.e., the position above the event). Further, each event satisfies $\Schemastocks$.
\end{example}

The notion of a \emph{renaming of a event} will be useful in this work (e.g., see Section~\ref{sec:acea}). Formally, we define a \emph{renaming} $r$ as a mapping $r: \aset \mapsto \aset$.
We say that an event $e$ is consistent with a renaming $r$ iff $r(\ba) = r(\bb)$ then $e(\ba) = e(\bb)$.
Given an event $e$ consistent with $r$, we define the \emph{renamed} event $r(e)$ such that $[r(e)](\ctime) = e(\ctime)$ and $[r(e)](r(\ba)) = e(\ba)$ for every attribute $\ba \in \dom(r)$. In other words, $r$ renames each attribute $\ba$ of $e$ to $r(\ba)$.
We define by $\TR$ the set of all renamings over~$\aset$.

\paragraph{Predicates of events} A \emph{predicate} is a possibly infinite set $P$ of events. 
For instance, $P$ could be the set of all events $e$ such that $e(\ba) \leq \text{20}$.
In our examples, we will use the notation $\ba \sim a$ where $\ba \in \aset$, $a \in \dset$,  and $\sim$ is a binary relation over $\dset$ to denote the predicate $P = \{e \mid e(\ba) \sim a\}$.
We say that an event $e$ satisfies predicate $P$, denoted $e \models P$, if, and only if, $e \in P$.
We generalize this notation from events to a bag of events $E$ such that $E \models P$ if, and only if, $e \models P$ for every~$e \in E$. 

In this work, we assume a fix \emph{set of predicates} $\pset$ that is close under intersection,  negation, and renaming, namely, $P_1 \cap P_2 \in \pset$, $\eset \setminus P \in \pset$, and $r(P) \in \pset$ for every $P, P_1, P_2 \in \pset$ and $r \in \TR$ where $r(P) = \{r(e) \mid e \in P \, \wedge\,  e \text{ is consistent with } r\}$ and $\eset$, the set of all events, is a predicate in $\pset$ that we usually denote by $\texttt{TRUE}$.

\paragraph{Complex events} In this work, we will use a slightly different definition of complex event: we will store events inside valuations, instead of storing positions like in~\cite{BucchiGQRV22}. %
Formally, fix a finite set $\xset$ of \emph{variables}, which includes all event types (i.e. $\tset \subseteq \xset$). Let $\Stream$ be a stream of length~$n$. A \emph{complex event}
of $\Stream$ is a triple $(i, j, \mu)$ where $i, j \in [n]$, $i \leq j$, and $\mu:\xset \rightarrow \bagpow{\eset}$ is a function from variables to 
finite bags of events. Intuitively, $i$ and $j$ marks the beginning and end of the interval where the complex event happens, and $\mu$ stores the events in the interval $[i..j]$ that fired the complex event.
In the following, we will usually denote $C$ to denote a complex event $(i, j, \mu)$ of $\Stream$ and omit $\Stream$ if the stream is clear from the context. We will use $\ctime(C)$, $\cstart(C)$, and $\cend(C)$ to denote the interval $[i..j]$, the start $i$, and the end $j$ of $C$, respectively. Further, by some abuse of notation we will also use $C(X)$ for $X \in \xset$ to denote the bag $\mu(X)$ of $C$.  

The following operations on complex events will be useful throughout the paper. We define the \emph{union} of complex events $C_1$ and $C_2$, denoted by $C_1 \uplus C_2$, as the complex event $C'$ such that $\cstart(C') = \min\{\cstart(C_1), \cstart(C_2)\}$, $\cend(C') = \max\{\cend(C_1), \cend(C_2)\}$, and $C'(X) = C_1(X) \uplus C_2(X)$ for every $X \in \xset$. Further, we define the \emph{projection over $L \subseteq \xset$} of a complex event $C$, denoted by $\pi_L(C)$, as the complex event $C'$ such that $\ctime(C') = \ctime(C)$ and $C'(X) = C(X)$ whenever $X \in L$, and $C'(X) = \emptyset$, otherwise. Finally, we denote by $(i,j, \mu_\emptyset)$ the complex event with the trivial function $\mu_\emptyset$ such that $\mu_\emptyset(X) = \emptyset$ for every $X \in \xset$.

\paragraph{A new semantics for CEL}
In this work, we use the \emph{Complex Event Logic} (\cel) introduced in~\cite{GrezRUV21} and implemented in CORE~\cite{BucchiGQRV22} as our basic query language for CER. However, we revisit its semantics in order to extend it with aggregation. In particular, we use the same \cel syntax as in~\cite{GrezRUV21} which is given by the following grammar:
\[
\begin{array}{rcllcll}
	\varphi & := &  R & \text{(event type selection)}
	& \mid & \varphi \kAS X & \text{(variable binding)} \\ 
	& \mid & \varphi \kFILTER X[P] & \text{(predicate filtering)} 
	& \mid & \kPROJ_L(\varphi) & \text{(variable projection)} \\
	& \mid & \varphi \kOR \varphi & \text{(disjunction)}  
	& \mid & \varphi \kAND \varphi  & \text{(conjunction)} \\
	& \mid & \varphi \kSSEQ \varphi & \text{(contiguous sequencing)}  
	& \mid & \varphi \kSEQ \varphi & \text{(non-cont. sequencing)} \\
	& \mid & \varphi\kSITER & \text{(contiguous iteration)} 
	& \mid & \varphi\kITER & \text{(non-cont. iteration)}
\end{array}
\]
where $R$ is an event type, $X \in \xset$ is a variable, $P \in \pset$ is a predicate, and $L \subseteq \xset$ is a finite set of variables.  
Similar to~\cite{GrezRUV21,BucchiGQRV22}, we define the semantics of a \cel formula $\varphi$ over a stream $\Stream = e_1 e_2 \ldots e_n$, recursively, as a set of complex events over $\Stream$. The main difference is the notion of complex events, that now contains events instead of positions. In Figure~\ref{fig:cel-semantics}, we define the semantics of each \cel operator like in~\cite{BucchiGQRV22,GrezRUV21}. Given a formula $\varphi$, the semantics $\sem{\varphi}(S)$ defines a set of complex events. Notice that $\sem{\varphi}(S)$ has a \emph{set-semantics} and, instead, complex events store bags of events.

\begin{figure}[t]
	\small
	\centering
	\begin{align*}
		\sem{R}(\Stream) & \ = \ \{ (i,i, \mu) \ \mid \!\!
		\begin{array}[t]{l}
			i \in [k] \, \wedge \, e_i(\type) = R \, \wedge \, \mu(R) = \bleft e_i\bright \, \wedge \,  \forall Y \neq X. \, \mu(Y) = \emptyset\} 
		\end{array}  \\ 
		\sem{\varphi \kAS X}(\Stream) & \  = \  \{ C \ \mid \!\!
		\begin{array}[t]{l}
			\exists \, C' \in \sem{\varphi}(\Stream). \ \ \ctime(C) = \ctime(C') \ \wedge \ C(X) = \biguplus_{Y} C'(Y)   \\
			\wedge \ \forall Z \neq X. \  C(Z) = C'(Z) \} 
		\end{array} \\
		\sem{\varphi \kFILTER X[P]}(\Stream) & \ = \  \{ C \ \mid \!\!
		\begin{array}[t]{l}
			C \in \sem{\varphi}(\Stream)  \ \wedge \ C(X) \models P \} 
		\end{array} \\
		\sem{\kPROJ_{L} (\varphi)}(\Stream) & \ = \ \{ \kPROJ_L(C) \ \mid \!\!
		\begin{array}[t]{l}
			C \in \sem{\varphi}(\Stream) \}
		\end{array} \\
		\sem{\varphi_1 \kOR \varphi_2}(\Stream) & \ = \ \sem{\varphi_1}(\Stream) \ \cup \  \sem{\varphi_2}(\Stream) \\
		\sem{\varphi_1 \kAND \varphi_2}(\Stream) & \ = \ \sem{\varphi_1}(\Stream) \ \cap \  \sem{\varphi_2}(\Stream) \\
		\sem{\varphi_1 \!\kSSEQ\! \varphi_2}(\Stream) & \ = \ \{ C_1 \uplus C_2 \ \mid \!\!
		\begin{array}[t]{l}
			C_1 \in \sem{\varphi_1}(\Stream) \, \wedge \,  C_2 \in \sem{\varphi_2}(\Stream) \, \wedge \, \cend(C_1) + 1 = \cstart(C_2) \}
		\end{array}  \\
		\sem{\varphi_1 \kSEQ \varphi_2}(\Stream) & \ = \ \{ C_1 \uplus C_2 \ \mid \!\!
		\begin{array}[t]{l}
			C_1 \in \sem{\varphi_1}(\Stream) \, \wedge \, C_2 \in \sem{\varphi_2}(\Stream) \, \wedge \, \cend(C_1) < \cstart(C_2)   \}
		\end{array}  \\
		\sem{\varphi\kSITER}(\Stream) & \ = \ \sem{\varphi}(\Stream) \ \cup \ \sem{\varphi \ssq \varphi \kSITER}(\Stream)  \\
		\sem{\varphi\kITER}(\Stream) & \ = \ \sem{\varphi}(\Stream) \ \cup \ \sem{\varphi \sq \varphi \kITER}(\Stream) 
	\end{align*}
	 \vspace{-5mm}
	\caption{The semantics of \cel defined over a stream $\Stream = e_1 e_2 \ldots e_n$ where each $e_i$ is an event.}
		\label{fig:cel-semantics}
\end{figure}

Next, we present an example for showing how to use the syntax and semantics of \cel to extract complex events from streams (see also Section~\ref{sec:aggregation}). In this example, we use conjunction and disjunction in filtering that one can read them as:
\[
\begin{array}{rcl}
	\varphi \kFILTER (X[P_1] \wedge Y[P_2]) & \equiv & (\varphi \kFILTER X[P_1]) \kFILTER Y[P_2] \\
	\varphi \kFILTER (X[P_1] \vee Y[P_2]) & \equiv & (\varphi \kFILTER X[P_1]) \kOR (\varphi \kFILTER Y[P_2])
\end{array}
\]
for every \cel formula $\varphi$, variables $X, Y \in \xset$, and predicates $P_1, P_2$.

\begin{example}[from \cite{BucchiGQRV22}]\label{ex:stocks-basic}
	Consider the stream $\Streamstocks$ from Example~\ref{ex:stream}.  Suppose that we are interested in all triples of \texttt{SELL} events where the first is a sale of Microsoft over US\$100, the second is a sale of Intel (of any price), and the third is a sale of Amazon below US\$2000. Then, we can specify this pattern by the following CEL formula:
	\[ 
	\begin{array}{rcl}
		\varphi_2 & \!\!= \!\! & \big(\texttt{SELL} \kAS \operatorname{msft} \kSEQ \texttt{SELL} \kAS \operatorname{intel} \kSEQ \texttt{SELL} \kAS \operatorname{amzn}\big) \\
		& & \ \ \ \ \   \kFILTER \big(\mathsf{msft[name = ``MSFT"]} \ \wedge \ \mathsf{msft[price > 100]} \ \wedge \ \mathsf{intel[name = ``INTC"]} \\
		& & \qquad \qquad \qquad \wedge \ \mathsf{amzn[name = ``AMZN"]} \ \wedge \ \mathsf{amzn[price < 2000]}\big).
	\end{array}	
	\]
	Intuitively, the expression $(\texttt{SELL} \kAS \operatorname{msft} \kSEQ \texttt{SELL} \kAS \operatorname{intel} \kSEQ \texttt{SELL} \kAS \operatorname{amzn})$ specifies that we want to see three \texttt{SELL} events that we named by the variables $\operatorname{msft}$, $\operatorname{intel}$ and $\operatorname{amzn}$, respectively. The semicolon operator ($\kSEQ$) indicates non-contiguous sequencing among them, namely, there could be more events between them. Finally, the $\kFILTER$ clause requires the data of the events to satisfy the necessary restrictions.
\end{example}

As we already mentioned, in this work we change the semantics used in \cite{GrezRUV21,BucchiGQRV22} to use events instead of positions, called it here \emph{event-based} semantics. The old semantics of \cel, called \emph{position-based} semantics, was obtained by outputting complex events of the form $C = (i, j, \mu_{\operatorname{index}})$ where $\mu_{\operatorname{index}}$ a mapping such that $\mu_{\operatorname{index}}: \xset \mapsto \pow{\bbN}$. Namely, $\mu_{\operatorname{index}}$ contains the positions of the events that participates in $C$. One can easily see that the event-based semantics of \cel is equivalent to the position-based semantics where $i$-th position must be replaced by the $i$-th event of the stream. In other words, we have the following equivalence.
\begin{theorem}\label{theo:oldvsnew}
	The (old) position-based semantics of \cel is equivalent to the (new) event-based semantics of \cel.
\end{theorem}

\paragraph{A new operator for projecting attributes}
An advantage of providing a new event-based semantics is that one can extend \cel with new operators, such as aggregation, which we will discuss in the next chapters. 
More interestingly, we can introduce new natural operators for managing complex events that cannot be defined using the old semantics in~\cite{GrezRUV21}. In this work, we use events instead of positions, which makes it possible to extend the \cel syntax with the \emph{attribute-projection operator}, an operator for projecting tuples within complex events. Formally, we extend the syntax of \cel formulas with the following operator:
\[
\begin{array}{rcll}
	\varphi & := & \kPROJ_{X(\ba_1, \ldots, \ba_k)}(\varphi) \ \ \ \  & \text{(tuple projection)} 
\end{array}
\]
where $\varphi$ is an arbitrary \cel formula, $X$ is a variable in $\xset$, and $\ba_1, \ldots, \ba_k$ is a list of attributes in $\aset$. Intuitively, it means that it will only consider the attributes in $\ba_1, \ldots, \ba_k$ in the events that are in the variable $X$.

We define the formal semantics of the attribute-projection operator $\kPROJ_{X(\ba)}$ recursively as follows. For a list of attributes $\ba_1, \ldots, \ba_k$ and an event $e$, we define $\pi_{\ba_1, \ldots, \ba_k}(e)$ as the new event $e'$ such that $\att(e') = \att(e) \cap \{\ba_1, \ldots, \ba_k\}$, $e(\ctime) = e'(\ctime)$, and $e'(\ba_i) = e(\ba_i)$ whenever $\ba_i \in \att(e')$. 
Let $\Stream = e_1 e_2 \ldots e_n$ where each $e_i$ is an event. Then:
\begin{align*}
	\sem{\kPROJ_{X(\ba_1, \ldots, \ba_k)}\!(\varphi)}(\Stream) & = \big\{\, C \ \mid \!\!
		\begin{array}[t]{l}
			\exists C' \in \sem{\varphi}(\Stream). \ \ctime(C) = \ctime(C') \, \wedge \, \forall Y \neq X. \, C(Y) = C(Y') \\ \wedge \, C(X) = \bleft \pi_{\ba_1, \ldots, \ba_k}(e) \mid e \in C'(X) \bright \big\}
		\end{array}
\end{align*}
Intuitively, given a complex event $C' \in \sem{\varphi}(\Stream)$ with $C'(X) = \bleft e_1, \dots , e_l\bright$, the projection formula above creates a new complex event $C$ which has the same interval than $C'$ and events in variables $Y \neq X$, but it redefines events in $X$ as $C(X) = \bleft  \pi_{\ba_1, \ldots, \ba_k}(e_1), \dots ,  \pi_{\ba_1, \ldots, \ba_k}(e_l)\bright$.

\begin{example}
	We consider again the setting as in Examples~\ref{ex:stream} and~\ref{ex:stocks-basic}. Now we are interested in getting the price of the sale of Intel, subject to the same constraints. Then, we can write this query by using tuple projection as follows:
	\[ 
	\begin{array}{rcl}
		\varphi_4 & \!\!= \!\! & \kPROJ_{\operatorname{intel}(\texttt{price})}\big(\texttt{SELL} \kAS \operatorname{msft} \kSEQ \texttt{SELL} \kAS \operatorname{intel} \kSEQ \texttt{SELL} \kAS \operatorname{amzn}\big) \\
		& & \ \ \ \ \   \kFILTER \big(\mathsf{msft[name = ``MSFT"]} \ \wedge \ \mathsf{msft[price > 100]} \ \wedge \ \mathsf{intel[name = ``INTC"]} \\
		& & \qquad \qquad \qquad \wedge \ \mathsf{amzn[name = ``AMZN"]} \ \wedge \ \mathsf{amzn[price < 2000]}\big).
	\end{array}	
	\]
\end{example} 	
	\section{Modelling aggregate functions in CER}\label{sec:aggregation-func}

Before introducing our logic for aggregation, we present a framework to model aggregate functions in CER based on monoids. Our goal is to present a logic that is as general as possible, encompassing most of the aggregations queries used in practice, such as $\sumtext$, $\max$, $\cnt$, or $\rng$. In the following, we recall the definitions of monoids and aggregate functions. We end by stating our main assumptions regarding aggregation in CER.

\paragraph{Monoids} 
A \emph{monoid} is an algebraic structure $(M, \oplus, \zero)$  where  $(M, \oplus)$ forms a semigroup and $\zero\in M$ is an identity element over $\oplus$. Similar to semigroups, we will further assume that $\oplus$ is commutative. For example, the natural numbers with addition $(\bbN, +, 0)$ forms a commutative monoid and the natural numbers without zero and product $(\bbN \setminus \{0\}, \times, 1)$ also forms a commutative monoid. Other examples are $(\bbN \cup \{\infty\}, \min, \infty)$ with $\min$ and $(\bbN, \max, 0)$ with $\max$. 
Given a commutative monoid $(M, \oplus, \zero)$, a finite bag $A = \bleft a_1, \ldots, a_n \bright \subseteq M$ and a function $f:M \rightarrow M$ we define the operator: $\bigoplus_{a \in A} f(a) = f(a_1) \oplus \ldots \oplus f(a_n)$, namely, the generalization of $\oplus$ from a binary operator to a set of elements. In particular, if $A = \emptyset$, we define~$\bigoplus_{a \in A} f(a) = \zero$. In the sequel, we will use $(M,\oplus, \zero)$ or $(M,\otimes,\one)$ for denoting arbitrary commutative monoid over some set $M$. 

\paragraph{Aggregate functions} 
In this work, we consider the most general definition of an aggregate function that can be defined through a monoid (see  \cite{jesus2014survey}). Specifically, an \emph{aggregate function} is a function from a bag of values to values, formally,
$
f: \bagpow{\dset} \rightarrow \dset
$.
An aggregate function $f$ is \emph{self-decomposable} if there exists a commutative monoid\footnote{In~\cite{jesus2014survey}, the definition of self-decomposable is not given in terms of a monoid. However, one can easily see that the definition in~\cite{jesus2014survey} implies the existence of a monoid.}
$(M, \oplus, \zero)$ such that $f(X \uplus Y) = f(X) \oplus f(Y)$ for every disjoint bags $X, Y \subseteq \dset$.
Examples of functions that are self-decomposable are $\sumtext$, $\max$, $\min$, and $\cnt$. For instance, 
$\sumtext (X \uplus Y)$ is equal to $0$ if $X \uplus Y = \emptyset$, to $x$ if $X \uplus Y = \bleft x \bright$, and to $\sumtext (X) + \sumtext (Y)$, otherwise. Similarly, $\cnt (X \uplus Y)$ is equal to $0$ if $X \uplus Y = \emptyset$, to $1$ if $X \uplus Y = \bleft x \bright$, and to $\cnt (X) + \cnt (Y)$ otherwise.
Finally, $\min (X \uplus Y)$ is equal to	$\infty$ if $X \uplus Y = \emptyset$, to $x$ if $X \uplus Y = \bleft x \bright$, and $\min(\min (X),\min (Y))$~otherwise.

Unfortunately, in practice not all aggregate function are self-decomposable; however, most of them can still be decomposed before we apply a simple operation. Formally, an aggregate function $f$ is called \emph{decomposable} if there exist a function $g: M \rightarrow M'$ and a self-decomposable aggregate function $h$ such that 
$
f = g \circ h
$.
Furthermore, we assume that $g$ can be computed in constant time (i.e., in the RAM model). This last condition is necessary, as we want $g$ to perform a simple operation (i.e., constant time), as the final step after $h$ has completed the aggregation, and not to be powerful enough to perform the aggregation itself.

Every self-decomposable functions is also decomposable (i.e., where $g$ is the identity function). 
Other examples of aggregate functions that are decomposable (but not self-decomposable) are $\avg$ and $\rng$. For instance, one can define $\avg$ as $\avg(X) = g(h(X))$ where $h(\bleft x \bright) = (x,1)$ and $h(X \uplus Y)= h(X)+h(Y)$ 
where $+$ is the standard pointwise sum of pairs, and $g((s,c)) = s/c$. Another example is the $\rng$ which can be defined as $\rng(X) = g(h(X))$ such that $h({x}) =(x,x)$, $h(X \uplus Y) = (\max(X \uplus Y), \min(X \uplus Y))$, and $g((s,c)) = s-c$. In both cases, one can check that $h$ is a self-decomposable function and $g$ can be computed in constant time in the RAM model.

Notice that, although self-decomposable functions can be decomposed through a monoid, they are not entirely specified by it (e.g., $\cnt$). Nevertheless, as the following lemma shows, we can restrict to monoids by first mapping the values to the underlying monoid. 
\begin{lemma}\label{lemma:self-decomp}
	$f$ is self-decomposable if, and only if, there exist a commutative monoid $(M, \oplus, \zero)$ and a function $f': \dset \rightarrow M$ such that $f(X) = \bigoplus_{a \in X} f'(a)$ for every bag $X$.
\end{lemma}

Given the previous lemma, we say that $f$ is \emph{strong} self-decomposable if there exists a pair $(M, f')$ such that $M$ is a commutative monoid and $f'$ is the identity function. In other words, $f$ can be directly defined by a commutative monoid. The functions that are strong self-decomposable are $\sumtext$, $\min$, and $\max$. On the other hand, $\cnt$ needs to map each value to $1$ before adding them. 

For the sake of simplification, in the following we assume that all aggregate functions are \emph{strong self-decomposable}. In other words, we can directly define the semantics of the aggregate functions through a commutative monoids. We can make this assumption since the functions $f'$ and $g$ (i.e., of decomposable aggregate functions) can be computed in constant time when the each data item is read or after the aggregation is done, like, for example, the function $f'$ to map a single element to 1 (e.g., $\cnt$), and the final function $g$ to calculate the difference between two elements (e.g., $\rng$), or divide one by the other (e.g., $\avg$). This assumption considerably simplifies our setting, allowing us to focus on the most relevant details of aggregation without discarding relevant aggregate functions from practice.%
 	
	\section{Aggregation complex event logic}\label{sec:aggregation}

In this section, we present our proposal to extend \cel with aggregation. Specifically, we demonstrate how to extend \cel with an operation for aggregations, building upon previous work experience. We provide examples of how this new operator is sufficient to model most queries used in earlier works. We start by introducing the algebraic structure for modelling aggregate functions, which we then use to define the aggregation operator in \cel.

\paragraph{The algebraic structure for aggregation} Recall that $\aset$ and $\dset$ are our fix sets of attributes names and data values, respectively. We fix an algebraic structure:
\[
\cD \ = \  (\dset, \oplus_1, \ldots, \oplus_k, \zero_1, \ldots, \zero_k) \tag{$\dagger$}
\]
over $\dset$ such that each $(\dset, \oplus_i, \zero_i)$ forms a commutative monoid for every $i \in [k]$. For example, $(\bbN \cup \{\infty\}, +, \min, \max, \zero, \infty, \zero)$ forms such an algebraic structure where we assume that $n + \infty = \infty$ for every $n \in \bbN$. Without loss of generality, we assume that $\nullvalue \in \dset$ and $a \oplus_i \nullvalue = \nullvalue$ for every $a \in \dset$ and $i \in [k]$ (if this is not the case, one can extend $\cD$ with a fresh value $\nullvalue$). The purpose of $\nullvalue$ is to define the aggregation operator over events $e$ where an attribute is not defined (i.e., $e(\ba) = \nullvalue$ for some $\ba\in \aset$).

\paragraph{The single-attribute aggregation operator} Our goal is to extend the syntax of \cel with an \emph{aggregation operator} that aggregates values in a single event. For the sake of presentation, we will first introduce the operation for a single attribute to then show how to extend it to multiple attributes. 

Specifically, we extend the \cel syntax with the \emph{single-attribute aggregation operator}, called \emph{Aggregation \cel} (\acel), as follows:
\[
\varphi \ := \ \agg_{Y[\bb \leftarrow \otimes X(\ba)]}(\varphi)
\]
where $\varphi$ is an arbitrary \cel formula, $X$ and $Y$ are variables in $\xset$, $\ba$ and $\bb$ are attributes names in $\aset$, and $\otimes$ is a  binary operator from $\cD$ where $(\dset, \otimes, \one)$ forms a monoid.
Intuitively, the syntax $Y[\bb \leftarrow \otimes X(\ba)]$ means that the aggregation will create a new event $e$ that will be stored at the variable $Y$, such that $e$ will have a single attribute $\bb$ that stores the $\otimes$-aggregation of the $\ba$-attribute of events in $X$.  
We define the formal semantics of the single aggregation operator $\agg$ recursively as follows. Let $\Stream = e_1 e_2 \ldots e_n$ be a stream. Then:
\begin{multline*}
	\sem{\agg_{Y[\bb \leftarrow \otimes X(\ba)]}(\varphi)}(\Stream) \ = \\
	\big\{ C \ \mid \!
	\begin{array}[t]{l}
		\exists C' \in \sem{\varphi}(\Stream). \ \ctime(C) = \ctime(C') \, \wedge \, \forall Z \neq Y. \  C(Z) = C'(Z) \\
		\wedge \ C(Y) = C'(Y) \uplus \bleft e \mid e = [\bb \mapsto \bigotimes\limits_{e' \in C'(X)} e'(\ba)] \wedge e(\ctime) = \cend(C') \bright \ \big\}
	\end{array}
\end{multline*}
Intuitively, given a complex event $C' \in \sem{\varphi}(\Stream)$ with 
$C'(X) = \bleft e_1, \ldots, e_\ell\bright$, the aggregation formula above creates a new complex event $C$ which has the same interval and events than $C'$ except that $Y$ has an additional event $e$ (i.e., 
$C(Y) = C(Y') \uplus \bleft e\bright$) 
and $e(\bb) = e_1(\ba) \otimes \cdots  \otimes e_\ell(\ba)$. In case that $C'(X) = \emptyset$, it will return the identity $\one$ of $\otimes$.
Further, the new event $e$ has $e(\ctime) = \cend(C')$, namely, the last time inside $C'$. Notice that the event $e$ is always well-defined, since we assume that, if $\ba$ is not defined for some~$e_i$, it holds that $e_i(\ba) = \nullvalue$ and then $e(\ba) = \nullvalue$.

In the following, we use some special notation for useful functions like $\sumtext$, $\max$, $\min$ instead of~$\oplus$. For example, if we use the sum function, we write $\agg_{Y[\bb \leftarrow \sumtext(X(\ba))]}(\varphi)$. Further, recall that, although we use commutative monoids to define the semantics, this semantics can easily be generalized to \emph{decomposable aggregate functions} like $\cnt$, $\avg$, or $\rng$.
Therefore, without loss of generality, we also write, for example, $\agg_{Y[\bb \leftarrow \cnt(X(\ba))]}(\varphi)$ although strictly speaking $\cnt$ is not a strong self-decomposable aggregate function.%

\begin{example}\label{ex:stocks-agg1}
	Consider again the setting as in Examples~\ref{ex:stream} and \ref{ex:stocks-basic}. Now we are interested in getting the maximum price in a sequence of Intel sales between a Microsoft and an Amazon sale under the same constrains and store it in an attribute $\texttt{MAX}$ in a variable $M$. Then, we can specify this query by using the aggregation operator as:
	\[ 
	\begin{array}{rcl}
		\varphi_6 & = & \agg_{M[\texttt{MAX} \gets \max(\operatorname{intel}(\texttt{price}))} \big( \\
		& & \qquad [\texttt{SELL} \kAS \operatorname{msft} \kSEQ (\texttt{SELL} \kAS \operatorname{intel}) \kITER \kSEQ \texttt{SELL} \kAS \operatorname{amzn}] \\
		& & \qquad \kFILTER [\mathsf{msft[name = ``MSFT"]} \ \wedge \ \mathsf{msft[price > 100]} \ \wedge \ \mathsf{intel[name = ``INTC"]} \\
		& & \qquad \qquad  \qquad \wedge \ \mathsf{amzn[name = ``AMZN"]} \ \wedge \ \mathsf{amzn[price < 2000]}] \big)
	\end{array}	
	\]
	As the reader can check from the semantics of $\agg$, the max value of the intel sequence will be stored in a new event at the variable $M$.
\end{example}

\begin{example}\label{ex:stocks-agg2}
	For a second example, suppose that now we are interested in getting the length of a sequence $(\texttt{BUY} \kOR \texttt{SELL})$ in an upward trend between prices 100 and 2000 and store it in an attribute $\texttt{QNT}$ in a variable $Q$. Further, we want to also check that this length is greater than $5$. Then, we can express the query~as:
	\[ 
	\begin{array}{rcl}
		\varphi_7 & = & \Big[\agg_{M[\texttt{QNT} \gets \cnt(\operatorname{m}(\texttt{price}))} \big(\\
		& & \qquad [(\texttt{BUY} \kOR \texttt{SELL}) \kAS l \kSEQ (\texttt{BUY} \kOR \texttt{SELL}) \kITER \kAS m \kSEQ (\texttt{BUY} \kOR \texttt{SELL}) \kAS h] \\
		& & \qquad \qquad \kFILTER [\mathsf{l[price < 100]} \ \wedge \ \mathsf{m[price \geq 100]}\\
		& & \qquad \qquad \qquad \wedge \ \mathsf{m[price \leq 2000]} \ \wedge \ \mathsf{h[price > 2000]}] \big) \Big]  \kFILTER \ \mathsf{M[QNT > 5]}
	\end{array}	
	\]
\end{example}
Notice that, although in the previous example the aggregation was applied over a simple \cel formula (i.e., at the topmost level), in \acel all operators, including the aggregation operator, can be freely composed. In particular, we can apply a filter (e.g. $\kFILTER \ \mathsf{M.QNT > 5}$) over an aggregation that was computed.

\paragraph{The multi-attribute aggregation operator} We present now the generalization of the aggregation operator to multiple attributes. Although this generalized version is more verbose, it is needed in practice for aggregating different sets simultaneously in different attributes. 
We extend the syntax of \cel with the \emph{(multi-attribute) aggregation operator}:
\[
\varphi \ := \ \agg_{Y[\bb_1 \leftarrow \otimes_1 X_1(\ba_1), \ldots, \bb_\ell \leftarrow \otimes_\ell X_\ell(\ba_\ell)]}(\varphi)
\]
where $\varphi$ is an arbitrary \cel formula, $X_1, \ldots, X_\ell$ and $Y$ are variables in $\xset$, $\ba_1, \ldots, \ba_\ell$ and $\bb_1, \ldots, \bb_\ell$ are attributes names in $\aset$, and $\otimes_1, \dots, \otimes_\ell$ are binary operators from $\cD$.
Intuitively, the syntax $Y[\bb_1 \leftarrow \otimes_1 X_1(\ba_1), \ldots, \bb_\ell \leftarrow \otimes_\ell X_\ell(\ba_\ell)]$
states that the aggregation will create a new event $e$ with attributes $\bb_1, \ldots, \bb_\ell$ that will be stored at the variable $Y$, such that each attribute $\bb_i$ will store the $\otimes_i$-aggregation of the $\ba_i$-attribute of events in $X_i$.

Given a stream $\Stream$, the formal semantics of the generalization of $\agg$ is given as follows:
\begin{multline*}
	\sem{\agg_{Y[\bb_1 \leftarrow \otimes_1 X_1(\ba_1), \ldots, \bb_\ell \leftarrow \otimes_\ell X_\ell(\ba_\ell)]}(\varphi)}(\Stream) \ = \\ \big\{ C \! \mid \!\!
	\begin{array}[t]{l}
		\exists C' \in \sem{\varphi}(\Stream). \, \ctime(C) = \ctime(C') \, \wedge \,
        \forall Z \neq Y. \,  C(Z) = C'(Z) \, \wedge \,  C(Y) = C'(Y) \, \uplus \\
		\bleft e \mid e = [\bb_1 \mapsto 
		{\sideset{}{_1}\bigotimes\limits_{e' \in C'(X_1)}} e'(\ba_1), \ldots, \bb_\ell \mapsto{\sideset{}{_\ell}\bigotimes\limits_{e' \in C'(X_\ell)}} e'(\ba_\ell)] \wedge e(\ctime) = \cend(C')\bright \! \big\}
	\end{array}
\end{multline*}
Intuitively, the general version of $\agg$ allows to define several attributes $\bb_1, \ldots, \bb_\ell$ by performing aggregation over the attributes $\ba_1, \ldots, \ba_\ell$, respectively. The idea is similar to the single-attribute aggregation operator but with several attributes $\bb_1, \ldots, \bb_\ell$ at once.

In Appendix~\ref{sec:all-examples}, we show how to use \acel to specify several examples from previous academic proposals and real-life systems. In particular, we present examples from the literature where the multi-attribute aggregation operator is required. Similar to the simple-attribute aggregation operator, in \acel, one can freely compose all operators, including this new aggregation operator. We conclude this section by discussing several relevant design decisions we made in defining the aggregation operator in \cel.

\paragraph{Why this semantics for aggregation in \cel?}
There are multiple ways to define a semantics for aggregation in CEL; however, our proposal for CEL and ACEL has some crucial design decisions that need to be justified. Specifically, we propose a semantics that (1) outputs a set of complex events (i.e., no repetitions), (2) each complex event contains bags of events, and (3) each event has a timestamp that defines the time when it arrives or was created. Indeed, we could consider other alternatives, such as a semantics that outputs bags of complex events, sets inside a complex event, or events without a timestamp, or any combination of these alternatives. In the following, we discuss why we proposed a semantics based on (1), (2), and (3), and what the consequences are of taking other alternatives.

For (1), if we choose a semantics based on bags of complex events, independent of the other choices, we will get a semantics that outputs duplicated results depending on how we specify the query. For example, assume a bag-based semantics and a user writes the query:
\[
\varphi_1 \ = \ \kPROJ_X (A \kAS X \kSEQ B \kITER \kSEQ A \kAS X)
\] 
over a stream $\Stream_1 = A_1B_2B_3A_4$ where $A$ and $B$ are the types of the events (i.e., the data in the attributes is not relevant). If we evaluate $\varphi_1$ over $\Stream_1$ with a bag-based semantics, we will have the same result $(1,4)$ multiple times (potentially exponentially many times) depending on how many $B$ were captured for each result. 
Instead, a set-based semantics ensures that each complex event appears only once, no matter how the query is specified. 

For (2), if we choose that each variable inside a complex event maps to a set of events, instead of a bag of events, we could get some answers that do not consider some results as they will be taken as repeated elements. For example, if we consider a query: %
\[
\varphi_2 \ = \ \agg_{Y (\bb \gets \sumtext(X(\ba)))} \big[(\agg_{X(\ba \gets \sumtext(A(\valvar)))} [B \kSSEQ A \kSITER]) \kSITER\big]
\]
and a stream $\Stream_2 = B_1A_2[\ba: 3]A_3[\ba: 5]B_4A_5[\ba: 2]A_6[\ba: 4]A_7[\ba: 2]$, first we will get two matches (one from $B_1A_2A_3$ and the other from $B_4A_5A_6A_7$), then we will make the aggregation in each of them, but the result of each aggregation is the same (i.e., $8$), they come from different values and they will not be saved as two different values,
so finally the outer aggregation will be applied over one element and not two.

Finally, for (3), if we consider that each event that arrives or is created does not have a timestamp (i.e., a mark of origin), then we can still lose some information during aggregation (even if we used bags inside complex events to store events). For example, consider the query 
\[
\varphi_3 \ = \  (\agg_{Y (\bb \gets \sumtext (X(\ba)))}(X \kSITER)\kSEQ \psi_1) \kOR (\psi_2 \kSEQ \agg_{Y (\bb \gets \sumtext (W(\ba)))}(W \kSITER))
\]
for some subformulas $\psi_1$ and $\psi_2$. 
For formula $\varphi_3$ over some stream, the results of the aggregation in the left and right parts of the disjunction (i.e., $\kOR$) could be equal, and it will be impossible to  differentiate which part the aggregation is coming from when we apply the $\kOR$ operator. Instead, by assuming that each event has a timestamp (even those created through aggregation), for $\varphi_3$, there will be at least two outputs, and we can differentiate the position where the aggregation was performed.

It is important to note that another semantics for \cel and the aggregation operator is possible, and our argument above does not invalidate them. However, there could be consequences for the query language with unintuitive behavior for the users. In this work, we have chosen to focus on a semantics based on (1), (2), and (3), studying its properties, and reserve the study of other variants of the logic for future work.

\paragraph{Expressive power of \acel}
When introducing a new operator, such as aggregation, one wants it to model only what it is meant to; however, combining it with other operators can lead to unexpected properties that can be expressed. In particular, combining the aggregate operator with filters is very powerful, as it allows one to check equivalence between events. For example, consider the following query with aggregation and filtering:
\[
\kPROJ_{X,Y} \big( \agg_{Z(\bb_1 \gets \sumtext (X(\ba)), \bb_2 \gets \sumtext(Y(\ba)))} (R \kAS X \kSEQ T \kAS Y) \kFILTER [Z (\bb_1 = \bb_2)] \big)
\]
Intuitively, the previous query checks that an event of type $T$ (naming it $Y$) happens after an event of type $R$ (naming it $X$) and $\sumtext$ the values of attribute $\ba$ in both events separately, saving those values in attributes $\bb_1$ and $\bb_2$ of variable $Z$ in one event. Then, the query filters it by checking if the values of attributes $\bb_1$ and $\bb_2$ are equal, as they correspond to the same event. Finally, it projects variables $X$ and $Y$. One can see that the query correlates events from different types only using aggregation and filter operators over single events. 

This unexpected behavior of combining aggregation with filtering is an interesting side effect that could lead to a better understanding of aggregation in CER. Note that observing this interaction between aggregates, filters, and other operators will not be possible without having a concrete and formal semantics of the query language. 	
	\section{Automata model for aggregation in CER} \label{sec:acea}

Here we present an automata model for aggregation that extends complex event automata with registers similar to the model of cost register automata~\cite{AlurDDRY13}. We start by recalling the model of complex event automata (\CEA) to provide then the necessary definitions for introducing our new automata model for aggregation.

\paragraph{CEA}
A \emph{Complex Event Automaton (\CEA)}~\cite{GrezRUV21,BucchiGQRV22} is a tuple
$
\cA = (Q, \Delta, q_0, F)
$
where $Q$ is a finite set of states, $\Delta \subseteq Q\times \pset \times \pow{\xset} \times Q$ is a finite transition relation, $q_0 \in Q$ is the initial state, and $F \subseteq Q$ is the set of final states. %
A \emph{run of $\cA$ over stream $\cS = e_1 \ldots e_n$ from positions $i$ to $j$} is a sequence of transition:
\[
\rho \ := \  q_{i} \ \trans{P_{i}/L_i} \ q_{i+1} \ \trans{P_{i+1}/L_{i+1}} \ \ldots \ \trans{P_j/L_j} \ q_{j+1}
\]
such that $q_i = q_0$ is the initial state of $\cA$ and for every $k \in [i..j]$  it holds that $(q_{k}, P_{k}, L_k, q_{k+1}) \in \Delta$ and $e_k \models P_k$. A run $\rho$  is \emph{accepting} if $q_{j+1} \in F$.
An accepting run $\rho$ of $\cA$ over $\Stream$ from $i$ to $j$ naturally defines the complex event
$
C_{\rho} \ := \ (i,j, \mu_\rho)
$
such that 
$\mu_\rho(X) = \{ t_k \mid i \leq k \leq j \wedge X \in L_k \}$ for every $X \in \xset$.
If position $i$ and $j$ are clear from the context, we say that $\rho$ is a run of $\cA$ over $\Stream$. 
Finally, we define the semantics of $\cA$ over a stream $\Stream$ as:
$
\semaux{\cA}(\Stream) \ := \ \{C_\rho \mid \text{$\rho$ is an accepting run of $\cA$ over $S$} \}.
$

CEA was crucial to capture the expressiveness of \cel, compile queries from \cel into CEA, and efficiently evaluate them. Unfortunately, one can easily notice that CEA are not useful for our \acel semantics, since there is no way to \emph{remember} the values of the attributes that we have seen to do the aggregation. In other words, there is no mechanism for aggregating values and producing new events as in the new semantics of \acel. 
We will show how to overcome these shortcomings in the next definitions.

\paragraph{Expressions} Recall that $\aset$ is a fixed set of attributes and $\dset$ a fix set of data values. Further, recall that in Section~\ref{sec:aggregation} we fix an algebraic structure of the form ($\dagger$)
over $\dset$ such that each $(\dset, \oplus_i, \zero_i)$ forms a commutative monoid for every $i \in [k]$.
We define an \emph{$(\cD, \aset)$-expression} $e$ (or just \emph{expression}) as a syntactical formula over $\cD$ and $\aset$ generated by the grammar: 
\[
    \alpha  \ := \ d \ \mid\  \ba \ \mid \ \alpha \oplus_i  \alpha \ \ \  \ i \in [k]
\]
where $d \in \dset$ and $\ba \in \aset$. We define the \emph{set of all $(\cD, \aset)$-expressions} by $\expr(\cD, \aset)$. 
For any expression $\alpha \in \expr(\cD, \aset)$ we denote by $\att(\alpha)$ the set of all attributes in $\alpha$. 
Given an event $e: \aset \mapsto \dset$ and an expression $\alpha$ such that $\att(\alpha) \subseteq \att(e)$, we define the semantics of $\alpha$ over $e$, denoted by $\sem{\alpha}(e)$,  as the value in $\dset$ of evaluating $\alpha$ by replacing every $\ba \in \aset$ by~$e(\ba)$.
\begin{example}\label{ex:expressions}
	   Let $\cD =  (\dset, \min, \max, +, \infty, 0, 0)$, the expressions $\alpha = \ba + \bb$ and $\beta = \min(\ba,\bb) + \max(\bb,\bc)$, and an event $e$ where $e(\ba) = 4$, $e(\bb) = 2$ and $e(\bc) = 5$. Then the result of each expression $\alpha$ and $\beta$ over the event $e$ are $\sem{\alpha}(e) = 6$ and $\sem{\beta}(e) = 7$, respectively.
\end{example}

\paragraph{Assignments} An \emph{$(\cD, \aset)$-assignment} (or just \emph{assignment} when $\cD$ and $\aset$ are clear from the context) is a program that assigns attributes in $\aset$ to expressions in $\expr(\cD, \aset)$. Formally, an \emph{assignment} is defined as a mapping 
$\sigma: \aset \mapsto \expr(\cD, \aset)$. 
Similar to expressions, we define
$\attin(\sigma) = \bigcup_{\ba \in \dom(\sigma)} \att(\sigma(\ba))$ to be all the attributes used in expressions of $\sigma$, and $\attout(\sigma) = \dom(\sigma)$ all the attributes that are assigned.
Given an event $e: \aset \rightarrow \dset$ and an $(\cD, \aset)$-assignment $\sigma$ such that $\attin(\sigma) \subseteq \att(e)$, the semantics of an assignment $\sigma$ over $e$ is an event $e':= \sem{\sigma}(e): \aset \mapsto \dset$ such that $\att(e') = \attout(\sigma)$ and  $e'(\ba) = \sem{\sigma(\ba)}(e)$ for every $\ba \in \attout(\sigma)$. 
In other words, $\sem{\sigma}(e)$ is the result of applying the assignment $\sigma$ with the values in the event $e$. 
We denote the set of all $(\cD, \aset)$-assignments by $\assigns(\cD,\aset)$. 

\begin{example}
    Consider again the setting of Example~\ref{ex:expressions} and the assignment $\sigma$ defined as:
    $
    \sigma : \ \ \ba \ \gets \ \max(\ba+\bb,\bc)
    $.
    Here, we think $\sigma$ as a program where the left side of $\gets$ is updated with the right side, namely, $\sigma(\ba) = \max(\ba+\bb,\bc)$. Then, $[\sem{\sigma}(e)](\ba) = 6$.%
\end{example}

Finally, we recall the notion of renamings (Section~\ref{sec:cel}) and define updates of events that will be useful for our automata model. So, remember that a \emph{renaming} $r$ is defined as $r: \aset \mapsto \aset$, which maps each attribute to a new attribute. We can note that a renaming is also a particular case of an assignment $r: \aset \mapsto \expr(\cD, \aset)$ such that $r(\ba) \in \aset$.
Also, recall that we define by $\TR$ the set of all tuple renaming over~$\aset$.
Given events $e$ and $e'$, we define the \emph{update of $e'$ by $e$}, denoted by $e  \gg e'$, as a new event such that $\att(e  \gg e') = \att(e) \cup \att(e')$ and $[e  \gg e'](\ba) = e(a)$ if $\ba \in \att(e)$, and $[e \gg e'](\ba) = e'(\ba)$ otherwise. %

\paragraph{Aggregation Complex Event Automata\label{subsec:acea}}
We are ready to define the model of CEA with aggregation. An \emph{Aggregation Complex Event Automaton (\ACEA)} is a tuple
$
\cA \ = \ (Q, \Delta, q_0, F)
$  
where $Q$ is a finite set of states, $q_0 \in Q$ is the initial state, $F \subseteq Q$ are the final states, and:
\[
\Delta \subseteq Q \times \assigns(\cD,\aset) \times \pset \times \{\lambda: \xset \mapsto \bagpow{\TR}\} \times Q
\]
is a finite transition relation where $\{\lambda: \xset \mapsto \bagpow{\TR}\}$ is the set of all mappings $\lambda$ that maps a variable $X$ to a finite bag of renamings $\bleft r_1, \ldots, r_k\bright$.
A transition $(p, \sigma, P, \lambda, q) \in \Delta$ specifies that $\cA$ can move from state $p$ to state $q$ after reading an event, by updating some internal registers with $\sigma$ and checking a condition (over the registers) with $P$. Similar to \CEA, $\lambda$ will be in charge of creating the outputs of the complex event where the renamings $\lambda(X) = \bleft r_1, \ldots, r_k\bright$ will create $k$ new tuples in the variable $X$ coming from the values stored in the~internal registers. 
We assume that the renamings in $\lambda$ for a transition of the form $(p, \sigma, P, \lambda, q) \in \Delta$ are consistent with $\sigma$, namely, $\attin(\lambda(X)) \subseteq \dom(\sigma)$ for every $X \in \dom(\lambda)$.

A pair $(q, \nu)$ is a configuration of $\cA$ where $q \in Q$ and $\nu:\aset \mapsto \dset$ is an event which represents the current values of the attributes.
For the sake of simplification, in \ACEA, we use attributes as ``registers'' for storing temporary values. For this reason, the configuration $(q, \nu)$ represents that the automata is in the state $q$ and the registers $\dom(\nu)$ (i.e., a subset of attributes) store the current computed values.

Let $\Stream = e_1 \ldots e_n$ be a stream. A \emph{run of $\cA$ over stream $S$ from positions $i$ to $j$} is a sequence of configurations and transitions: 
\[
\rho \ := \  (q_{i}, \nu_i) \, \lltrans{\sigma_i,P_{i}/\lambda_i} \, (q_{i+1}, \nu_{i+1}) \, \llltrans{\sigma_{i+1},P_{i+1}/\lambda_{i+1}} \, \ldots \, \lltrans{\sigma_j,P_j/\lambda_j} \, (q_{j+1}, \nu_{j+1}) \tag{$\ddagger$}
\]
such that $q_i$ is the initial state $q_0$, $\nu_i$ is the empty event (i.e., $\dom(\nu_i) = \emptyset$), and for every $k \in [i..j]$, $(q_k, \sigma_k,P_{k},\lambda_k, q_{k+1}) \in \Delta$, $(q_{k+1}, \nu_{k+1})$ is a configuration of $\cA$ with
$\nu_{k+1} := \sem{\sigma_{k}}(e_k \gg \nu_k)$, and $\nu_{k+1} \models P_{k}$.
Also, it must hold that $\attin(\sigma_k) \subseteq \dom(e_k \gg \nu_k)$.
Intuitively, the new values $\nu_{k+1}$ are produced by first updating $\nu_k$ by the new event $e_k$ (i.e., $e_k \gg \nu_k$) and then operate $e_k \gg \nu_k$ by the assignment $\sigma_k$. After the new values $\nu_{k+1}$ are computed, we check if they satisfy the predicate $P_k$ of the transition.

Similar to CEA, 
a run $\rho$ is \emph{accepting} if $q_{j+1} \in F$. An accepting run $\rho$ like ($\ddagger$) of $\cA$ over $\Stream$ from $i$ to $j$ defines the complex event $C_{\rho} := (i,j, \mu_\rho)$ such that:
\[
\mu_\rho(X) = \bleft e \mid  k \in [i..j] \wedge r \in \lambda_k(X)  \wedge e = \sem{r}(\nu_{k+1}) \wedge e(\ctime) = k \bright
\]
for every $X \in \xset$. 
Finally, we define the semantics of $\cA$ over a stream $\Stream$ as:
\[
\semaux{\cA}(\Stream) := \{ C_\rho \mid \text{$\rho$ is an accepting run of $\cA$ over $S$} \}.
\]

\begin{figure}[tbp]
	{
	\small
    \begin{center}
    \begin{tikzpicture}[->,>=stealth, semithick, auto, initial text= {}, initial distance= {3mm}, accepting distance= {4mm}, node distance=2.2cm, semithick]
        \tikzstyle{every state}=[draw=black,text=black,inner sep=4pt, minimum size=5mm]
            
            \begin{scope}
        \node[initial,state]    (1)         {$q_1$};
        \node[state] (2) [right=4.5cm of 1] {$q_2$};
        \node[state, accepting] (3) [right=4.55cm of 2]  {$q_3$};

        \path
        (1) edge node {$\left[\begin{array}{l}
        		\bm \leftarrow 0 \\
        		\bn \leftarrow \texttt{name} \\
        		\bp \leftarrow \texttt{price}
        	\end{array}\right], {P_1} \ \mid \ \lambda_1$} (2)
        (2) edge node {$\left[\begin{array}{l}
        		\bm \leftarrow \bm \\
        		\bn \leftarrow \texttt{name} \\
        		\bp \leftarrow \texttt{price}
        	\end{array}\right], P_3 \ \mid \ \lambda_3$} (3)
        (2) edge [loop below] node {$
        	\left[\begin{array}{l}
        		\bm \leftarrow \max\{\bm,\texttt{price}\} \\
        		\bn \leftarrow \texttt{name} \\
        		\bp \leftarrow \texttt{price}
        	\end{array}\right], P_2 \ \mid \ \lambda_2$} (2);

        \node at ($(1) + (-1,0.05)$) {$\cA$:};

        \end{scope}
            \end{tikzpicture}
            \vspace{-2mm}
        \end{center}
    }
    \caption{An \ACEA $\cA$ representing the given query in Example~\ref{ex:stocks-agg1} where $P_1 := \bn = \mathsf{``MSFT"} \wedge \bp > \mathsf{100}$, $P_2 := \bn = \mathsf{``INTC"}$ and $P_3 := \bn = \mathsf{``AMZN"} \wedge \bp < \mathsf{2000}$. Further, $\lambda_1(\mathsf{msft}) = \lambda_2(\mathsf{intel}) = \lambda_3(\mathsf{amzn}) = \bleft [\mathsf{name} \mapsto \bn, \mathsf{price} \mapsto \bp] \bright$, and $\lambda_3(\mathsf{M}) = \bleft [\mathsf{MAX} \mapsto \bm] \bright$.
    }
    \label{fig:ex-acea1}
\end{figure}

\begin{example}\label{ex:acea-1}
    Consider the \acel query from Example~\ref{ex:stocks-agg1}. We can obtain the same result with the \ACEA $\cA$ in Figure \ref{fig:ex-acea1} where $P_1 := \bn = \mathsf{``MSFT"} \wedge \bp > \mathsf{100}$, $P_2 := \bn = \mathsf{``INTC"}$ and $P_3 := \bn = \mathsf{``AMZN"} \wedge \bp < \mathsf{2000}$. Further, $\lambda_1(\mathsf{msft}) = \lambda_2(\mathsf{intel}) = \lambda_3(\mathsf{amzn}) = \bleft [\mathsf{name} \mapsto \bn, \mathsf{price} \mapsto \bp] \bright$, and $\lambda_3(\mathsf{M}) = \bleft [\mathsf{MAX} \mapsto \bm] \bright$. Intuitively, in the first transition, $\cA$ initializes a register $\bm$ (i.e., an attribute) with $0$ and checks that the price and name attributes satisfy the predicate $P_1$, by storing the name in $\bn$ and the price in $\bp$. Then, in the loop of $q_2$, $\cA$ updates the maximum value in $\bm$ with the new price and again checks that the name satisfies~$P_2$. Finally, in the last transition, it maintains the maximum value in $\bm$ and verifies that the attributes name and price satisfy~$P_3$. The mappings $\lambda_1$, $\lambda_2$, and $\lambda_3$ are in charge of outputting the events in variables $\mathsf{msft}$, $\mathsf{intel}$, and $\mathsf{amzn}$, respectively. Further, $\lambda_3$ is in charge of producing the final event in variable $M$ that contains the max-aggregate of Intel's prices. 
\end{example}

A first natural question to answer is whether the expressive power of the new model \ACEA includes queries defined by CEA or not. Similar to the question of \acel versus \cel, CEA outputs complex events with positions, where our new model outputs complex events with events among other new features. Below, we show that a \ACEA can define every \CEA by mapping the positions of the stream to the events. 

\begin{theorem}\label{th:cea-in-acea}
 \ACEA can define the same as \CEA over streams over a schema $\Sigma$, namely, for every \CEA $\cA$ there exists an \ACEA $\cA'$ such that $\sem{\cA}(\Stream) = \sem{\cA'}(\Stream)$ for every $\Stream$ over $\Sigma$.
\end{theorem}

\paragraph{Equivalence with ACEL\label{sec:eq-acel-acea}}
The first main goal of this paper is to provide a query language with a formal and denotational semantics for performing aggregation in CER. The second main goal is to provide a computational model to compile queries from this language. In the following result, we show that \ACEA is a computational model to fulfill this goal. Specifically, we show that every formula $\varphi$ in \acel can be compiled into a \ACEA, proving that the model has all the feature to perform complex event extraction and aggregation.

\begin{theorem}\label{th:acel-in-acea}
    Let $\Sigma$ be a schema. For every \acel formula~$\varphi$, there exists an \ACEA $\cA_\varphi$ such that $\sem{\varphi}(\Stream) = \sem{\cA_\varphi}(\Stream)$ for every stream $\Stream$ over~$\Sigma$.
\end{theorem}

We present the proof in Appendix~\ref{sec:proofs-acea}. It goes by induction over the formula showing how to compile each operator into an \ACEA. The standard \cel operators follow a similar construction to that in \cite{GrezRUV21} (except the AND operator), but here we also have to make sure that the registers are correctly maintained to produce the output. %

It is important to remark that \ACEA is a \emph{hybrid automata model} that needs to perform computation (i.e., for the aggregation), check filters (i.e., for the predicates), and produce outputs (i.e., events). Therefore, in designing the model, we seek an equilibrium that fulfills all these goals and, simultaneously, is as simple as possible. This simplicity could be helpful for understanding its expressiveness and designing efficient evaluation algorithms.

Despite its simplicity, \ACEA has more expressive power than \acel, namely, there are queries that can be defined with \ACEA but not with \acel. For example, consider the monoid of natural numbers $(\bbN, +, 0)$ (i.e., $\sumtext$). Given a stream $R[\ba: 1] \ \overset{n\text{-times}}{\ldots} \ R[\ba:1]$, one can define an \ACEA with one register that always doubles the current value and outputs its content in an event $[\bb: 2^n]$. Intuitively, \acel with $(\bbN, +, 0)$ cannot specify this query since it can only produce values that grow linearly with respect to the sum of all values in the stream. Even if we restrict the use of registers in a copyless manner (see \emph{copyless cost register automaton} in~\cite{AlurDDRY13}), one can design \ACEA that cannot be specified by \acel. For instance, given the previous stream, one can code an \ACEA that produces a complex event with the sequence of events: $[\bb: 1]  [\bb: 2] \ldots [\bb: n]$ (i.e., by adding in a register the input values and outputting its content in each transition). Given that in \acel, each value of an event can contribute to a finite number of new events, one cannot specify this in \acel. Therefore, \ACEA is more expressive than \acel, and it is an interesting open problem to characterize \acel in terms of restrictions over \ACEA. We leave this problem for future work.

	\section{Future Work}\label{sec:conclusions}

This paper provides logical foundations for aggregation in CER but leaves several open problems for future work. 
One relevant open problem is to better understand the equivalence between \acel and \ACEA, namely, which \ACEA can be written in \acel.
Another interesting question is to understand the expressive power of aggregation combined with filters and other operators (see Section~\ref{sec:aggregation}).
Finally, a crucial line of research for making \acel work in practice is to study how to evaluate \acel queries efficiently, finding enumeration algorithms that, given an \ACEA and a stream, run with \emph{constant update time and constant delay enumeration}.

	\bibliography{biblio}
	
	\newpage
	\appendix
	
	\section{Proofs from Section $\textbf{\ref{sec:cel}}$}\label{sec:proof-new-cel}

\subsection{Proof of Theorem~\ref{theo:oldvsnew}}
\begin{proof}
Let $\Stream$ be a stream and $\varphi$ be a query. Let $C = (i, j,\mu) \in \sem{\varphi} (S)$ where $C$ is obtained from the valuation semantics and each $\mu(X_i)$ is a set of positions corresponding to the events in variable $X_i$ (old semantics). On the other hand, let $C_{new} = (i,j,\mu_{new}) \in \sem{\varphi} (S)$ where $\mu$ is a mapping from variables to bags of events (new semantics). We can construct a function $f$ from the old semantics to the new such that $f : \mu(X) \rightarrow \mu_{new}(X)$, i.e., it takes each position in $\mu(X_i)$ to an event in $\mu_{new}(X_i)$. We can see that $f$ is injective and surjective because $\mu(X_i)$ and $\mu_{new}(X_i)$ come from they corresponding valuation of $\varphi$ and both semantics obtain the same events but one marks its position and the other the event in that position. 

\begin{figure}[tpbh]
	\small
	\begin{align*}
		\auxsem{R}(S) & = \{ (i,i,R \mapsto \{j\}) \mid S[j] \in \tuples(R) \}, \\
		\auxsem{\varphi \kAS A}(S) & = \{ (i,j,\mu[A \mapsto \sup(\mu)]) \mid \mu \in \auxsem{\varphi}(S) \}, \\
		\auxsem{\varphi \kFILTER P(A_1, \ldots, A_n)}(S) & = \{  (i,j,\mu \in \auxsem{\varphi}(S)) \mid (S[\mu(A_1)], \ldots, S[\mu(A_n)]) \in P \},\\
		\auxsem{\varphi_1 \kOR \varphi_2}(S) & = \auxsem{\varphi_1}(S) \cup \auxsem{\varphi_2}(S) \\
		\auxsem{\varphi_1 \kAND \varphi_2}(S) & = \auxsem{\varphi_1}(S) \cap \auxsem{\varphi_2}(S) \\
		\auxsem{\varphi_1 \kSEQ \varphi_2}(S) & = \{  (i,j,\mu_1 \cup \mu_2) \ \mid \!\!
		\begin{array}[t]{l} 
			\exists k, i \leq k < j: \mu_1 \in \auxsem{\varphi_1}(S,i,k),\\  
			\mu_2 \in \auxsem{\varphi_2}(S,k+1,j)\}, 
		\end{array}\\
		\auxsem{\varphi_1 \kSSEQ \varphi_2}(S) & = \{  (i,j,\mu_1 \cup \mu_2) \ \mid \!\!
		\begin{array}[t]{l} 
			\exists k, i \leq k < j: \mu_1 \in \auxsem{\varphi_1}(S,i,k),\\
			\mu_2 \in \auxsem{\varphi_2}(S,k+1,j), 
		\end{array}\\
		& \phantom{= \{  \mu_1 \cup \mu_2 \mid }\ \max(\sup(\mu_1)) = k, \min(\sup(\mu_2)) = k+1\}, \\
		\auxsem{\varphi \kITER}(S) & = \auxsem{\varphi}(S) \cup \auxsem{\varphi \kSEQ \varphi\kITER}(S) \\
		\auxsem{\varphi \kSITER}(S) & = \auxsem{\varphi}(S) \cup \auxsem{\varphi \kSSEQ \varphi\kSITER}(S) \\
		\auxsem{\kPROJ_L(\varphi)}(S) & = \{ (i,j,\mu|_L) \mid \mu \in \auxsem{\varphi}(S)\} \\
        \end{align*}
	\caption{The old semantics of \cel formulas defined over a stream $S = e_1 e_2 \ldots e_n$ where each $e_i$ is an event, and between positions $i$ and $j$.}
	\label{fig:old-semantics}
	\vspace{-2mm}
\end{figure}

\end{proof}
 	
	\section{Proofs from Section $\textbf{\ref{sec:aggregation-func}}$}
	\label{sec:proof-agg-func}

\subsection{Proof of Lemma~\ref{lemma:self-decomp}}
\begin{proof}
    First, we assume that $f'$ and $\oplus$ can be computed in constant time. Then, if $f$ is self-decomposable, we know that there exists a commutative monoid $(M, \oplus, \zero)$, and for every disjoint bags $X = \bleft a_1^X, \dots, a_k^X \bright$ and $Y = \bleft a_1^Y, \dots, a_l^Y \bright$ is true that $f(X \uplus Y) = f(X) \oplus f(Y)$. Following the definition, we can separate each bag recursively until we get the function applied to one element and we can see from the examples that $f(\bleft x \bright)$ can be $x$ or $1$, so we can generalize that by saying that exists a function $f'$ such that $f(\bleft x \bright) = f'(x)$. So, 
    \begin{align*}
        f(X \uplus Y) & = f(X) \oplus f(Y) \\ 
        & = f'(a_1^X) \oplus \dots \oplus f'(a_k^X) \oplus f'(a_1^Y) \oplus \dots \oplus f'(a_l^Y) \\
        & = \bigoplus_{a_i^X \in X} f'(a_i^X) \oplus \bigoplus_{a_i^Y \in Y} f'(a_i^Y) \\
        & = \bigoplus_{a \in X \uplus Y} f'(a)
    \end{align*}
    Finally, the condition is proved.

    Let $(M, \oplus, \zero)$ be a monoid, $f$ and $f': \dset \rightarrow M$ be functions such that for every bag~$X$, $f(X) = \bigoplus_{a \in X} f'(a)$. On the other hand, let $X = \bleft a_1^X, \dots, a_k^X \bright$ and $Y = \bleft a_1^Y, \dots, a_l^Y \bright$ be two disjoints bags, such that $f(X) = \bigoplus_{a_i^X \in X} f'(a_i^X)$ and $f(Y) = \bigoplus_{a_i^Y \in Y} f'(a_i^Y)$. We can expand $\bigoplus_{a_i^X \in X} f'(a_i^X)$ to $f'(a_1^X) \oplus \dots \oplus f'(a_k^X)$. Then, 
    \begin{align*}
        f(X \uplus Y) & = \bigoplus_{a \in X \uplus Y} f'(a) \\ 
        & = f'(a_1^X) \oplus \dots \oplus f'(a_k^X) \oplus f'(a_1^Y) \oplus \dots \oplus f'(a_l^Y) \\
        & = \bigoplus_{a_i^X \in X} f'(a_i^X) \oplus \bigoplus_{a_i^Y \in Y} f'(a_i^Y) \\
        & = f(X) \oplus f(Y)
    \end{align*}
    Finally, $f$ is self-decomposable.
\end{proof} 	
	\section{Proofs from Section $\textbf{\ref{sec:acea}}$}\label{sec:proofs-acea}

\subsection*{Proof of Theorem \ref{th:cea-in-acea}}
\begin{proof} 
    Let $\Sigma: \tset \rightarrow \pow{\asetSchema}$ be a schema. We assume that for each attribute $\ba \in \asetSchema$ there is a copy $\ba' \in \aset \setminus \asetSchema$. We will use this copy $\ba'$ to temporarily store values from the attributes in the register of the machine to produce identical copies of the events then.
    For a set $A \subseteq \asetSchema$, let $\bar{A} = \{\ba' \in \asetNotSchema \mid \ba \in A\}$.  We define $\sigma_{A}: A \cup \{\type\} \rightarrow \bar{A} \cup \{\type\}$ as the assignment that maps $\type$ to $\type$ and each $\ba \in A$ to its copy $\ba'$, namely, $\sigma_{A}(\type) = \type$ and  $\sigma_{A}(\ba) = \ba'$, otherwise. Note that $\sigma_{A}$ is a bijection and then $\sigma_A^{-1}$ is well-defined.

	Let $\cA = (Q, \Delta, q_0, F)$ be a \CEA. We define the same behavior of $\cA$ with an \ACEA as $\cA' = (Q, \Delta', q_0, F)$ over streams over $\Sigma$, where $Q$, $q_0$ and $F$ are the same and $\Delta'$ is define as:
    \begin{align*}
        \Delta' = \{(p, \sigma_{\Sigma(A)}, P_A, \lambda_A, 	q) \ \mid & \ (p, P, L, q) \in \Delta \ \wedge \ A \in \tset \\ 
        & \wedge \ \forall X \in L. \, \lambda_A(X) = \{[X \mapsto \{\sigma_{\Sigma(A)}^{-1}\}]\} \\ 
        & \wedge \ P_A = P \wedge (\type = A) \}
    \end{align*}
    This automaton has a transition for each $A \in \tset$ that updates its registers with its correspondent identity assignment, then checks that the predicate is for that $\type$, and finally uses its identity to create the tuple.

    We prove that $\sem{\cA}(\Stream) = \sem{\cA'}(\Stream)$ for a stream $\Stream$ over $\Sigma$. Let $C = (i, j, \mu_{\rho}) \in \sem{\cA}(\Stream)$ be a complex event over $\Stream = e_1 \ldots e_n$ and an accepting run of $\cA$ over $\Stream$ of the form:
    \[
    \rho \ := \  q_{i} \ \trans{P_{i}/L_i} \ q_{i+1} \ \trans{P_{i+1}/L_{i+1}} \ \ldots \ \trans{P_j/L_j} \ q_{j+1}
    \]
    such that $q_{j+1} \in F$ and $\mu_\rho(X) = \{ t_k \mid i \leq k \leq j \wedge X \in L_k \}$ for every $X \in \xset$.
	Then, by construction, we can find an accepting run of $\cA'$ over $\Stream$ of the form:
    \[
    \rho' := (q_i, \nu_i) \, \lltrans{\sigma_{i},P_i'/\lambda_i} \, (q_{i+1}, \nu_{i+1}) \, \trans{\dots} \, (q_{j}, \nu_{j}) \lltrans{\sigma_{j},P_j'/\lambda_j} \, (q_{j+1}, \nu_{j+1})
    \]
    where, for every $\ell \in \{i, \ldots, j\}$,  $\sigma_\ell = \sigma_{\Sigma(e_{\ell}(\type))}$, $P_\ell' =  P_\ell \wedge (\type = e_{\ell}(\type))$,  
    $\lambda_\ell(X) = \{[X \mapsto \{\sigma_{\Sigma(e_{\ell}(\type))}^{-1}\}]\}$ for every $X \in L_{\ell}$, and $\nu_\ell$ are defined accordingly. 
    Then, we can see that by definition of the transitions of $\cA'$ they have the same states as $\cA$, the same predicates but they also check the $\type$ of the event and instead of marking the events of the complex event with the set of variables in $L_k$, it uses renamings to mark the same events. Finally, we can see that $\mu_{\rho} = \mu_{\rho'}$ and $C \in \sem{\cA'}(\Stream)$.  
    
    One can easily check that the other direction follows by the same arguments.
\end{proof}

\subsection*{Proof of Theorem \ref{th:acel-in-acea}}
\begin{proof}
    We prove this result by constructing the \ACEA $\cA_\varphi = (Q, \Delta, q_0, F)$ by induction over the syntax of the formula $\varphi$ as follows. Note that we assume that this construction is for a specific schema $\Sigma: \tset \rightarrow \pow{\asetSchema}$ and we consider the same definition of $\sigma_A$ given in Theorem \ref{th:cea-in-acea}. 
    We define the empty assignment $\sigma_\varnothing$ as the assignment that independent of the input, it erase all the registers.
    We also assume that each automaton has a different set of registers, namely, if the automaton $\cA_{\psi_1}$ has the set of registers $A_1$ and $\cA_{\psi_2}$ has the set $A_2$, then $A_1 \cap A_2 = \varnothing$. 
    
    The proof goes by structural induction on an \acel formula $\varphi$. We start with the base case.
    
    \paragraph{[$\varphi = R$]} If $\varphi = R$, then $\cA_\varphi$ is defined as 
    \[
    \cA_\varphi = (\{p_1,p_2\},\{(p_1,\sigma_{\Sigma(R)},P_R,\underbrace{[R \mapsto \{\sigma_{\Sigma(R)}^{-1}\}]}_{\lambda},p_2)\},p_1,\{p_2\}),
    \]
    where $P_R$ is the predicate containing all tuples with type $R$, as was previously defined; and $\lambda$ is the function that for the variable $R$, it renames the attributes of the register tuple to its original, more specifically, $\lambda(R) = \{\sigma_{\Sigma(R)}^{-1}\}$. Intuitively, the automaton reads the event into its registers, checks that the type is $R$, and outputs the same events. We do this by storing the attributes into copies of attributes in $\Sigma(R)$. We require this condition to ensure that, if these attributes are used later in the construction, we will not overwrite them when new events arrive.
        
    \noindent We will prove that $\sem{\cA_R}(\Stream) = \sem{R}(\Stream)$. Let $\Stream = e_1 \dots e_n$ be a stream over the schema $\Sigma$.
    If $C \in \sem{\cA_R}(\Stream)$, then an accepting run over $\Stream$ is of the form: 
    \[
    \rho := (q_i, \nu_i) \, \lltrans{\sigma_{\Sigma(R)},P_R/\lambda} \, (q_{i+1}, \nu_{i+1})
    \]
    for some $R$ where $q_i = p_1$ and $q_{i+1} = p_2$. Then $C = (i, i, \mu)$ where $\mu = [R \mapsto \bleft e_i\bright)]$. We can note that it also satisfies the semantics defined before, so $C \in \sem{R}(\Stream)$. For the other direction, if $C \in \sem{R}(\Stream)$, then $C = (i, i, [R \mapsto \bleft e_i\bright])$. Given that $e_i(\type) = R$ and $\Stream$ satisfies the schema $\Sigma$, we know that $\att(e_i) = \Sigma(e_i(\type))$. 
    Then a run of the automaton over $\Stream$ from position $i$ to $i+1$ is $\rho := (p_1, \nu_1) \, \lltrans{\sigma_{\Sigma(R)},P_R/\lambda} \, (p_2, \nu_{2})$. Then, it is clear that $C \in \sem{\cA_R}(\Stream)$.
    
    We continue with the inductive cases. In the following, we will assume that for a formula $\psi$ we have an \ACEA $\cA_\psi$ that has the same output.
    
    \paragraph{[$\varphi = \psi \kAS X$]} If $\varphi = \psi \kAS X$, then $\cA_\varphi = (Q_\psi, \Delta_\varphi,q_{0\psi},F_\psi)$ where $\Delta_\varphi$ is the result of adding variable $X$ to all ``marking'' transitions of $\Delta_\psi$. Formally, 
    \[
    \begin{array}{rcl}
        \Delta_\varphi & = & \{(p,\sigma,P,\lambda,q)\in\Delta_\psi \mid \forall Y \in \xset. \ \lambda(Y) = \emptyset\} \\
        & \cup &  \{(p,\sigma,P,\lambda',q)\mid (p,\sigma,P,\lambda,q)\in \Delta_\psi \ \, \wedge \\
        & & \hspace{3cm} \lambda'(X) = \biguplus_{Z \in \xset} \lambda(Z)  \wedge   \forall Y \neq X. \ \lambda'(Y) = \lambda(Y)\}.
    \end{array}
    \]
    \noindent We will prove that $\sem{\cA_\varphi}(\Stream) = \sem{\psi \kAS X}(\Stream)$.
    If $C \in \sem{\cA_\varphi}(\Stream)$, then an accepting run over an stream $\Stream = e_1 \dots e_n$ is 
    \[
    \rho := (q_i, \nu_i) \, \lltrans{\sigma_{i},P_i/\lambda_i} \, (q_{i+1}, \nu_{i+1}) \, \trans{\dots} \, (q_{j}, \nu_{j}) \lltrans{\sigma_{j},P_j/\lambda_j} \, (q_{j+1}, \nu_{j+1})
    \]
    then $C = (i, j, \mu)$ where $\mu(X) = \biguplus_{Z \in \xset} \mu(Z)$. By definition $C' \in \sem{\cA_\psi}(\Stream)$ also has the same accepting run over the same $\Stream$, but it does not consider the variable $X$, being $C' = (i,j,\mu')$, and $\mu'(Y) = \mu(Y)$ for every $Y \neq X$. We can note that it also satisfies the semantics of the formula defined before, so $C \in \sem{\psi \kAS X} (\Stream)$.
    
    \noindent If $C \in \sem{\psi \kAS X} (\Stream)$ and $\Stream = e_1 \dots e_n$, then by definition:
    \[
    C = (i, j, \mu \cup [X \mapsto \biguplus_{Z \in \xset} \mu(Z)])
    \] and there exists $C' \in \sem{\psi}$ over the same stream such that $$C' = (i,j,\mu).$$ 
    By induction, a run of the automaton $\cA_{\psi}$ over $\Stream$ will be 
    \[
    \rho := (q_i, \nu_i) \, \lltrans{\sigma_{i},P_i/\lambda_i} \, (q_{i+1}, \nu_{i+1}) \, \trans{\dots} \, (q_{j}, \nu_{j}) \lltrans{\sigma_{j},P_j/\lambda_j} \, (q_{j+1}, \nu_{j+1})
    \]
    such that $C_\rho = (i, j, \mu) = C'$. Then, by construction:
    \[
    \rho' := (q_i, \nu_i) \, \lltrans{\sigma_{i},P_i/\lambda_i'} \, (q_{i+1}, \nu_{i+1}) \, \trans{\dots} \, (q_{j}, \nu_{j}) \lltrans{\sigma_{j},P_j/\lambda_j'} \, (q_{j+1}, \nu_{j+1})
    \]
    is a run of $\cA_{\varphi}$ and $C_{\rho'} = C$. We conclude that $C \in \sem{\cA_\varphi} (\Stream)$.

\paragraph{[$\varphi = \psi \kFILTER X\texttt{[}P\texttt{]}$]} If $\varphi = \psi \kFILTER X\texttt{[}P\texttt{]}$ for some variable $X$ and predicate $P$, then $\cA_\varphi = (Q_\psi, \Delta_\varphi,q_{0\psi},F_\psi)$ where $\Delta_\varphi$ is defined as:
\[
\begin{array}{rcl}
	\Delta_\varphi & = & \{(p,\sigma,P',\lambda,q)\in\Delta_\psi \mid \lambda(X) = \emptyset\} \\
	& \cup & \{(p,\sigma,P' \wedge \bigcap_{r \in \lambda(X)} r(P),\lambda,q)\mid (p,\sigma, P',\lambda,q)\in \Delta_\psi \wedge \lambda(X) \neq \emptyset\}.
\end{array}
\]
Assume that $\sem{\cA_{\psi}}(\Stream) = \sem{\psi}(\Stream)$. We will prove that: $$\sem{\cA_\varphi}(\Stream) = \sem{\psi \kFILTER X\texttt{[}P\texttt{]}}(\Stream).$$

\noindent Let $\Stream = e_1 \dots e_n$ be a stream. If $C \in \sem{\cA_\varphi}(\Stream)$, then an accepting run over $\Stream$ is of the form: 
\[
\rho := (q_i, \nu_i) \, \lltrans{\sigma_{i},P_i/\lambda_i} \, (q_{i+1}, \nu_{i+1}) \, \trans{\dots} \, (q_{j}, \nu_{j}) \lltrans{\sigma_{j},P_j/\lambda_j} \, (q_{j+1}, \nu_{j+1})
\]
where $C_\rho = (i,j, \mu) = C$. By construction, the automaton does the same as $\cA_\psi$, but the renamings in the transitions that marks the variable $X$ have to satisfy the predicate $P$, so every event in $\mu(X)$ satisfies $P$. Finally, it satisfies the definition of the formula and $C \in \sem{\psi \kFILTER X\texttt{[}P\texttt{]}}$.

\noindent In the other direction, if $C \in \sem{\varphi} (\Stream)$, then by definition $C \in \sem{\psi}(\Stream)$ and $C(X) \models P$. As $C \in \sem{\psi}(\Stream)$, then $C \in \sem{\cA_{\psi}}(\Stream)$. On the other hand, a run of $\cA_\varphi$ does the same as a run of $\cA_\psi$ but the events that are marked in variable $X$ also have to satisfy the predicate $P$ by definition. Finally, it satisfies the definition of the automaton and $C \in \sem{\cA_\varphi}(\Stream)$.

\paragraph{[$\varphi = \psi_1 \kOR \psi_2$]} If $\varphi = \psi_1 \kOR \psi_2$, then $\cA_\varphi$ is the union between $\cA_{\psi_1}$ and $\cA_{\psi_2}$, formally, 
\[
\cA_\varphi = (Q_{\psi_1} \cup Q_{\psi_2} \cup \{q_{0\varphi}\}, \Delta_{\varphi}, q_{0\varphi}, F_{\psi_1} \cup F_{\psi_2})
\]
where $q_{0\varphi}$ is a new state and $\Delta_{\varphi}  = \Delta_{\psi_1} \cup \Delta_{\psi_2} \cup \{(q_{0\varphi}, \sigma, P, \lambda, q) \mid (q_{0\psi_1}, \sigma, P, \lambda, q) \in \Delta_{\psi_1} \vee (q_{0\psi_2}, \sigma, P, \lambda, q) \in \Delta_{\psi_2}\}$. Here, we assume w.l.o.g. that $\cA_{\psi_1}$ and $\cA_{\psi_2}$ have disjoint sets of states.

\noindent We assume that $\sem{\cA_{\psi_1}}(\Stream) = \sem{\psi_1}(\Stream)$ and $\sem{\cA_{\psi_2}}(\Stream) = \sem{\psi_2}(\Stream)$, so we will prove that $\sem{\cA_\varphi}(\Stream) = \sem{\psi_1 \kOR \psi_2}(\Stream)$.
Let $C= (i, j, \mu) \in \sem{\cA_\varphi}(\Stream)$, then an accepting run over an stream $\Stream = e_1 \dots e_n$ is 
\[
\rho := (q_i, \nu_i) \, \lltrans{\sigma_{i},P_i/\lambda_i} \, (q_{i+1}, \nu_{i+1}) \, \trans{\dots} \, (q_{j}, \nu_{j}) \lltrans{\sigma_{j},P_j/\lambda_j} \, (q_{j+1}, \nu_{j+1})
\]
then by definition of the transitions, $\cA_\varphi$ can choose to execute $\cA_{\psi_1}$ or $\cA_{\psi_2}$, so $C \in \sem{\cA_{\psi_1}}(\Stream)$ or $C \in \sem{\cA_{\psi_2}}(\Stream)$. Finally, it satisfies the definition of the formula and $C \in \sem{\psi_1 \kOR \psi_2}(\Stream)$.

Let $C = (i, j, \mu) \in \sem{\psi_1 \kOR \psi_2}(\Stream)$ over a stream $\Stream$, so that means that $C \in \sem{\psi_1}(\Stream)$ or $C \in \sem{\psi_2}(\Stream)$. We know that if $C \in \sem{\psi_1}(\Stream)$ then $C \in \sem{\cA_{\psi_1}}(\Stream)$ (is analogous for $\psi_2$). Then, a run of the \ACEA $\cA_\varphi$ chooses if it runs $\cA_{\psi_1}$ or $\cA_{\psi_2}$, so the result $C_\varphi$ can be from $\cA_{\psi_1}$ or $\cA_{\psi_2}$, i.e., $C_\varphi \in \sem{\cA_{\psi_1}}(\Stream)$ or $C_\varphi \in \sem{\cA_{\psi_2}}(\Stream)$. Finally, it satisfies the definition of the automaton and $C \in \sem{\cA_\varphi}(\Stream)$.

\paragraph{[$\varphi = \psi_1 \kAND \psi_2$]} For the next construction we need some definitions. For two assignments $\sigma_1$ and $\sigma_2$ with $\attout(\sigma_1) \cap \attout(\sigma_2) = \emptyset$ we define the union $\sigma_1 \uplus \sigma_2$ as: 
$$\sigma_1 \uplus \sigma_2(\ba) = \begin{cases}
	\sigma_1 (\ba) \ \  \text{if } \ba \in \attout(\sigma_1) \\
	\sigma_2 (\ba) \ \  \text{if } \ba \in \attout(\sigma_2)
\end{cases}$$
for every $\ba \in \attout(\sigma_1) \cup \attout(\sigma_2)$.
For a bag of renamings $R$ we define its schema as the bag of sets of all the attributes in each renaming $r$:
\[
\schema(R) \ = \ \bleft \dom(r) \mid r \in R\bright.
\]

\noindent An isomorphism between bags of renamings $R_1$ and $R_2$ is a bijection $f: R_1 \rightarrow R_2$ such that for all renaming $r$ in $R_1$, $\dom(r) = \dom(f(r))$, implicitly, it means that the following holds $\schema(R_1) = \schema(R_2)$. Let $\ISO(R_1, R_2)$ be the set of all isomorphisms between $R_1$ and $R_2$.

\noindent We say that $\lambda_1$ is equivalent to $\lambda_2$, $\lambda_1 \equiv \lambda_2$, if there exists an isomorphism between them, $\dom(\lambda_1) = \dom(\lambda_2)$ and for all $X$ in $\xset$, $\schema(\lambda_1(X)) = \schema(\lambda_2(X))$. 

\noindent Coming back to the construction, if $\varphi = \psi_1 \kAND \psi_2$, then $\cA_\varphi$ is the intersection between $\cA_{\psi_1}$ and $\cA_{\psi_2}$, formally, $\cA_\varphi = (Q_{\psi_1} \times Q_{\psi_2}, \Delta_{\varphi}, (q_{0\psi_1}, q_{0\psi_2}), F_{\psi_1} \times F_{\psi_2})$, and 
\begin{align*}
	\qquad \Delta_{\varphi} = \{ & ((p_1, p_2), \sigma_1 \uplus \sigma_2, P_1 \wedge P_2 \wedge P_{\lambda_1, \lambda_2}, \lambda_1, (q_1, q_2)) \mid \\
	& (p_1, \sigma_1, P_1, \lambda_1, q_1) \in \Delta_{\psi_1} \ \wedge \ (p_2, \sigma_2, P_2, \lambda_2, q_2) \in \Delta_{\psi_2} \ \wedge \ \lambda_1 \equiv \lambda_2 \\
	& \wedge P_{\lambda_1, \lambda_2} = \bigwedge_{X \in \xset} \bigvee_{f \in \ISO(\lambda_1(X), \lambda_2(X))} \bigwedge_{r \in \lambda_1(X)} \bigwedge_{a \in \dom(r)} P_{r(a) = [f(r)](a)}\}
\end{align*}
We assume that w.l.o.g. that $\cA_{\psi_1}$ and $\cA_{\psi_2}$ have disjoint sets of assignments. Let  $\Stream = e_1 \dots e_n$ be a stream such that $\sem{\cA_{\psi_1}}(\Stream) = \sem{\psi_1}(\Stream)$ and $\sem{\cA_{\psi_2}}(\Stream) = \sem{\psi_2}(\Stream)$. So, we will prove that $\sem{\cA_\varphi}(\Stream) = \sem{\psi_1 \kAND \psi_2}(\Stream)$.

\noindent Let $C= (i, j, \mu) \in \sem{\cA_\varphi}(\Stream)$, then an accepting run over an stream $\Stream$ is of the form 
\[
\rho := (q_i, \nu_i) \, \lltrans{\sigma_{i},P_i/\lambda_i} \, (q_{i+1}, \nu_{i+1}) \, \trans{\dots} \, (q_{j}, \nu_{j}) \lltrans{\sigma_{j},P_j/\lambda_j} \, (q_{j+1}, \nu_{j+1})
\]
then by definition of the transitions, $\cA_\varphi$ executes $\cA_{\psi_1}$ and $\cA_{\psi_2}$ at the same time, maintaining both registers and checking both predicates, so $C \in \sem{\cA_{\psi_1}}(\Stream)$ and $C \in \sem{\cA_{\psi_2}}(\Stream)$. It also checks in the transition that for each variable $X$ there exists a bijection such that the renamings of $\lambda_1(X)$ have one that is equivalent in $\lambda_2(X)$, i.e., they write the same attribute with the same value. As the renamings are equivalent in the transition it only considers the ones from $\lambda_1$ for writing. Finally, it satisfies the definition of the formula and $C \in \sem{\psi_1 \kAND \psi_2}(\Stream)$.

\noindent Let $C = (i, j, \mu) \in \sem{\psi_1 \kAND \psi_2}(\Stream)$, so that means that $C \in \sem{\psi_1}(\Stream)$ and $C \in \sem{\psi_2}(\Stream)$. We know that if $C \in \sem{\psi_1}(\Stream)$ then $C \in \sem{\cA_{\psi_1}}(\Stream)$ (is analogous for $\psi_2$). Then, a run of the \ACEA $\cA_\varphi$ runs $\cA_{\psi_1}$ and $\cA_{\psi_2}$ simultaneously, so the result $C_\varphi$ is from $\cA_{\psi_1}$ and $\cA_{\psi_2}$, i.e., $C_\varphi \in \sem{\cA_{\psi_1}}(\Stream)$ and $C_\varphi \in \sem{\cA_{\psi_2}}(\Stream)$. Finally, it satisfies the definition of the automaton and $C \in \sem{\cA_\varphi}(\Stream)$.

\paragraph{[$\varphi = \psi_1 \kSEQ \psi_2$]} If $\varphi = \psi_1 \kSEQ \psi_2$, then $\cA_\varphi$ is the \ACEA that considers $\cA_{\psi_2}$ after $\cA_{\psi_1}$, formally, $\cA_\varphi = (Q_{\psi_1} \cup Q_{\psi_2}, \Delta_\varphi, q_{0\psi_1}, F_{\psi_2})$ where 
\begin{align*}
	\Delta_\varphi = & \Delta_{\psi_1} \cup \ \Delta_{\psi_2} \ \cup\ \{(q_{0\psi_2},\sigma_\varnothing,\TRUE,\emptyset,q_{0\psi_2}) \} \ \cup \\
	& \{(p,\sigma,P,\lambda,q_{0\psi_2}) \mid \exists q' \in F_{\psi_1}.(p,\sigma,P,\lambda,q') \in \Delta_{\psi_1}\}.
\end{align*}
Here, we assume w.l.o.g. that $\cA_{\psi_1}$ and $\cA_{\psi_2}$ have disjoint sets of states.

\noindent In the following, we assume that $\nu_\varnothing$ is the empty event, and that $\sem{\cA_{\psi_1}}(\Stream) = \sem{\psi_1}(\Stream)$ and $\sem{\cA_{\psi_2}}(\Stream) = \sem{\psi_2}(\Stream)$ over $\Stream = e_1 \ldots e_n$. So, we prove that $\sem{\cA_\varphi}(\Stream) = \sem{\psi_1 \kSEQ \psi_2}(\Stream)$.

\noindent Let $C= (i, j, \mu) \in \sem{\cA_\varphi}(\Stream)$, then an accepting run over $\Stream$ is of the form: 
\[
\rho := (q_i, \nu_i) \, \lltrans{\sigma_{i},P_i/\lambda_i} \, (q_{i+1}, \nu_{i+1}) \, \trans{\dots} \, (q_{j}, \nu_{j}) \lltrans{\sigma_{j},P_j/\lambda_j} \, (q_{j+1}, \nu_{j+1})
\]
then by definition of the transitions, $\cA_\varphi$ first executes $\cA_{\psi_1}$, then after it finishes it waits in the state $q_{0\psi_2}$ until it executes $\cA_{\psi_2}$, so we can separate the run as:
\[
\rho := \rho_1 \, \llltrans{\sigma_{{k+1}},P_{k+1}/\lambda_{k+1}} \, (q_{0\psi_2}, \nu_{\varnothing}) \, \trans{\dots} \, (q_{0\psi_2}, \nu_{\varnothing}) \llltrans{\sigma_{l-1},P_{l-1}/\lambda_{l-1}} \, \rho_2
\]
where $\rho_1$ and $\rho_2$ are of the form:
\[
\rho_1 := (q_i, \nu_i) \, \lltrans{\sigma_{i},P_i/\lambda_i} \, (q_{i+1}, \nu_{i+1}) \, \trans{\dots} \, (q_{k}, \nu_{k}) \lltrans{\sigma_{{k}},P_{k}/\lambda_{k}} \, (q_{k+1}, \nu_{k+1})
\]
\[
\rho_2 := (q_l, \nu_l) \, \lltrans{\sigma_{l},P_l/\lambda_l} \, (q_{l+1}, \nu_{l+1}) \, \trans{\dots} \, (q_{j}, \nu_{j}) \lltrans{\sigma_{j},P_j/\lambda_j} \, (q_{j+1}, \nu_{j+1})
\]
where $\rho_1$ is an accepting run of $\cA_{\psi_1}$ and $\rho_2$ is an accepting run of $\cA_{\psi_2}$, so we have $C_1 \in \sem{\cA_{\psi_1}}(\Stream)$ and $C_2 \in \sem{\cA_{\psi_2}}(\Stream)$. Then, $C_1 \in \sem{{\psi_1}}(\Stream)$, $C_2 \in \sem{{\psi_2}}(\Stream)$ and $C$ is the union of both results. Finally, it satisfies the definition of the formula and $C \in \sem{\psi_1 \kSEQ \psi_2}(\Stream)$.

\noindent Let $C = (i, j, \mu) \in \sem{\psi_1 \kSEQ \psi_2}(\Stream)$. This means that $C$ is the union of $C_1 \in \sem{\psi_1}(\Stream)$ and $C_2 \in \sem{\psi_2}(\Stream)$ and that $\cend(C_1) < \cstart(C_2)$. We know that if $C_1 \in \sem{\psi_1}(\Stream)$ then $C_1 \in \sem{\cA_{\psi_1}}(\Stream)$ (is analogous for $\psi_2$). Then, we can run $\cA_{\psi_1}$, wait in $q_{0\psi_2}$ and then execute $\cA_{\psi_2}$, to get the union of both complex events. Finally, it satisfies the definition of the \ACEA and $C \in \sem{\cA_\varphi}(\Stream)$.

\paragraph{[$\varphi = \psi_1 \kSSEQ \psi_2$]} If $\varphi = \psi_1 \kSSEQ \psi_2$, then $\cA_\varphi$ is the \ACEA that considers $\cA_{\psi_2}$ right after $\cA_{\psi_1}$, formally, $\cA_\varphi = (Q_{\psi_1} \cup Q_{\psi_2}, \Delta_\varphi, q_{0\psi_1}, F_{\psi_2})$ where 
\[ \qquad \Delta_\varphi = \Delta_{\psi_1} \cup \Delta_{\psi_2} \cup \{(p,\sigma,P,\lambda,q_{0\psi_2}) \mid (p,\sigma,P,\lambda,q_f) \in \Delta_{\psi_1} \wedge q_f \in F_{\psi_1}\}\]
Here, we assume w.l.o.g. that $\cA_{\psi_1}$ and $\cA_{\psi_2}$ have disjoint sets of states.

\noindent Similar to the previous construction, we assume that $\nu_\varnothing$ is the empty event, $\sem{\cA_{\psi_1}}(\Stream) = \sem{\psi_1}(\Stream)$ and $\sem{\cA_{\psi_2}}(\Stream) = \sem{\psi_2}(\Stream)$, so we will prove that $\sem{\cA_\varphi}(\Stream) = \sem{\psi_1 \kSSEQ \psi_2}(\Stream)$.

\noindent Let $C= (i, j, \mu) \in \sem{\cA_\varphi}(\Stream)$, then an accepting run over $\Stream$ is:
\[
\rho := (q_i, \nu_i) \, \lltrans{\sigma_{i},P_i/\lambda_i} \, (q_{i+1}, \nu_{i+1}) \, \trans{\dots} \, (q_{j}, \nu_{j}) \lltrans{\sigma_{j},P_j/\lambda_j} \, (q_{j+1}, \nu_{j+1})
\]
then by definition of the transitions, $\cA_\varphi$ first executes $\cA_{\psi_1}$, then right after it finishes, it executes $\cA_{\psi_2}$, so we can separate the run as:
\[
\rho := \rho_1 \, \llltrans{\sigma_{k},P_{k}/\lambda_{k}} \, \rho_2
\]
\[
\quad \rho_1 := (q_i, \nu_i) \, \lltrans{\sigma_{i},P_i/\lambda_i} \, (q_{i+1}, \nu_{i+1}) \, \trans{\dots} \, (q_{k-1}, \nu_{k-1}) \lltrans{\sigma_{k-1},P_{k-1}/\lambda_{k-1}} \, (q_{k}, \nu_{k})
\]
\[
\qquad \quad\rho_2 := (q_{k+1}, \nu_{k+1}) \, \lltrans{\sigma_{k+1},P_{k+1}/\lambda_{k+1}} \, (q_{k+2}, \nu_{k+2}) \, \trans{\dots} \, (q_{j}, \nu_{j}) \lltrans{\sigma_{j},P_j/\lambda_j} \, (q_{j+1}, \nu_{j+1})
\]
where $\rho_1$ with the transition $k$ is an accepting run of $\cA_{\psi_1}$ (if we consider that the transition $k$ in the original leads to a final state) and $\rho_2$ is an accepting run of $\cA_{\psi_2}$, so we have $C_1 \in \sem{\cA_{\psi_1}}(\Stream)$ and $C_2 \in \sem{\cA_{\psi_2}}(\Stream)$. Then, $C_1 \in \sem{{\psi_1}}(\Stream)$, $C_2 \in \sem{{\psi_2}}(\Stream)$ and $C$ is the union of both results. Finally, it satisfies the definition of the formula and $C \in \sem{\psi_1 \kSSEQ \psi_2}(\Stream)$.

\noindent Let $C = (i, j, \mu) \in \sem{\psi_1 \kSSEQ \psi_2}(\Stream)$, so that means that $C$ is the union of $C_1 \in \sem{\psi_1}(\Stream)$ and $C_2 \in \sem{\psi_2}(\Stream)$ and that $\cend(C_1) + 1 = \cstart(C_2)$. We know that if $C_1 \in \sem{\psi_1}(\Stream)$ then $C_1 \in \sem{\cA_{\psi_1}}(\Stream)$ (is analogous for $\psi_2$). Then, we can run $\cA_{\psi_1}$ and then right after execute $\cA_{\psi_2}$, connecting the last transition of $\cA_{\psi_1}$ to the first state of $\cA_{\psi_2}$, so we can get the union of both complex events. Finally, it satisfies the definition of the automaton and $C \in \sem{\cA_\varphi}(\Stream)$.

\paragraph{[$\varphi = \psi \kITER$]} If $\varphi = \psi \kITER$, then $\cA_\varphi = (Q_\psi \cup \{q_{new}\}, \Delta_\varphi, q_{0\psi}, F_\psi)$ where $q_{new}$ is a fresh state and $\Delta_\varphi$ is formally defined as follows:
\[
\begin{array}{rcl}
	\Delta_\varphi & = & \Delta_\psi \\
	& \cup &  \{(p,\sigma,P,\lambda,q_{new}) \mid \exists q' \in F_\psi.(p,\sigma,P,\lambda,q') \in \Delta_\psi\} \\
	& \cup & \{(q_{new},\sigma_\varnothing,\TRUE,\emptyset,q_{new})\} \\ 
	& \cup & \{(q_{new},\sigma,P,\lambda,p) \mid (q_{0\psi},\sigma,P,\lambda,p) \in \Delta_\psi\}.
\end{array}
\]
\noindent We assume that $\nu_\varnothing$ is the empty event. We will prove that $\sem{\cA_\varphi}(\Stream) = \sem{\psi \kITER}(\Stream)$.

\noindent Let $C= (i, j, \mu) \in \sem{\cA_\varphi}(\Stream)$, then an accepting run over $\Stream$ is of the form:
\[
\rho := (q_i, \nu_i) \, \lltrans{\sigma_{i},P_i/\lambda_i} \, (q_{i+1}, \nu_{i+1}) \, \trans{\dots} \, (q_{j}, \nu_{j}) \lltrans{\sigma_{j},P_j/\lambda_j} \, (q_{j+1}, \nu_{j+1})
\]
then by definition of the transitions, $\cA_\varphi$ executes $\cA_{\psi}$, then after it finishes it waits in the state $q_{new}$ until it executes $\cA_{\psi}$ again, so we can separate the run as:
\[
\rho := \rho'_1 \, \trans{\dots} \, (q_{new}, \nu_{\varnothing}) \, \trans{\dots} \, \rho'_2 \, \trans{\dots} \, (q_{new}, \nu_{\varnothing}) \trans{\dots} \, \rho'_l
\]
where each $\rho'_k$ is an accepting run of $\cA_{\psi}$, with $C'_k \in \sem{\cA_{\psi}}(\Stream)$ and therefore $C'_k \in \sem{{\psi}}(\Stream)$. Then, $C$ is union of the result of every run of $\cA_{\psi}$ that was done. Finally, it satisfies the definition of the formula and $C \in \sem{\psi \kITER}(\Stream)$.

\noindent Let $C = (i, j, \mu) \in \sem{\psi \kITER}(\Stream)$ over a stream $\Stream$, that means that $C$ is the union of complex events of having done the query $\psi$ multiple times, one after the other. We know that if each $C'_k \in \sem{\psi}(\Stream)$ then $C_k \in \sem{\cA_{\psi}}(\Stream)$. Then, we can run $\cA_{\psi}$, wait in $q_{new}$ and then execute $\cA_{\psi}$ again, multiple times, to get the union of all the complex events in each $C'_k$. Finally, it satisfies the definition of the automaton and $C \in \sem{\cA_\varphi}(\Stream)$.

\paragraph{[$\varphi = \psi \kSITER$]} If $\varphi = \psi \kSITER$, then we define $\cA_\varphi = (Q_\psi \cup \{q_{new}\}, \Delta_\varphi, q_{0\psi}, F_\psi)$ where $q_{new}$ is a fresh state and the transition relation $\Delta$ is defined as:
\[
\begin{array}{rcl}
	\Delta_\varphi & = & \Delta_\psi \\ 
	& \cup &  \{(p,\sigma,P,\lambda,q_{new}) \mid \exists q' \in F_\psi.(p,\sigma,P,\lambda,q') \in \Delta_\psi\} \\
	& \cup & \{(q_{new},\sigma,P,\lambda,p) \mid (q_{0\psi},\sigma,P,\lambda,p) \in \Delta_\psi\}.
\end{array}
\]

\noindent Next, we will prove that $\sem{\cA_\varphi}(\Stream) = \sem{\psi \kSITER}(\Stream)$ over a stream $\Stream = e_1 \ldots e_n$.

\noindent Let $C= (i, j, \mu) \in \sem{\cA_\varphi}(\Stream)$, then an accepting run over $\Stream$ is of the form:
\[
\rho := (q_i, \nu_i) \, \lltrans{\sigma_{i},P_i/\lambda_i} \, (q_{i+1}, \nu_{i+1}) \, \trans{\dots} \, (q_{j}, \nu_{j}) \lltrans{\sigma_{j},P_j/\lambda_j} \, (q_{j+1}, \nu_{j+1})
\]
then by definition of the transitions, $\cA_\varphi$ executes $\cA_{\psi}$, then after it finishes it executes $\cA_{\psi}$ again, so we can separate the run as:
\[
\rho := \rho'_1 \, \trans{\dots} \, \rho'_2 \, \trans{\dots} \, \dots \trans{\dots} \, \rho'_l
\]
where each $\rho'_k$ is an accepting run of $\cA_{\psi}$, with $C'_k \in \sem{\cA_{\psi}}(\Stream)$ and therefore $C'_k \in \sem{{\psi}}(\Stream)$. Then, $C$ is union of the result of every run of $\cA_{\psi}$ that was done just after the previous one. Finally, it satisfies the definition of the formula and $C \in \sem{\psi \kSITER}(\Stream)$.

\noindent Let $C = (i, j, \mu) \in \sem{\psi \kSITER}(\Stream)$, that means that $C$ is the union of complex events of having done the query $\psi$ multiple times, one just after the other. We know that if each $C'_k \in \sem{\psi}(\Stream)$ then $C_k \in \sem{\cA_{\psi}}(\Stream)$. Then, we can run $\cA_{\psi}$ and then $\cA_{\psi}$ again, multiple times, to get the union of all the complex events in each $C'_k$. Finally, it satisfies the definition of the automaton and $C \in \sem{\cA_\varphi}(\Stream)$.

\paragraph{[$\varphi = \kPROJ_L(\psi)$]} If $\varphi = \kPROJ_L(\psi)$ for some $L \subseteq \xset$, then $\cA_\varphi = (Q_\psi, \Delta_\varphi,q_{0\psi},F_\psi)$ where $\Delta_\varphi$ is the result of just consider the renamings that are associated with a variable in $L$ of each transition in $\Delta_\psi$. Formally, 
\[
\begin{array}{rcl}
	\Delta_\varphi & = & \{(p,\sigma,P, \lambda', q) \ \mid \ (p,\sigma,P,\lambda,q) \in \Delta_\psi \ \wedge  \\
	& & \hspace{3.1cm} \forall X \in L. \ \lambda'(X) = \lambda(X)  \ \wedge \ \forall X \notin L. \ \lambda'(X) = \emptyset\}
\end{array}
\]
\noindent Fix a stream $\Stream = e_1 \ldots e_n$ and assume that $\sem{\cA_{\psi}}(\Stream) = \sem{\psi}(\Stream)$. In the following, we will prove that $\sem{\cA_\varphi}(\Stream) = \sem{\kPROJ_L(\psi)}(\Stream)$.

\noindent Let $C \in \sem{\cA_\varphi}(\Stream)$, then an accepting run over $\Stream$ is of the form: 
\[
\rho := (q_i, \nu_i) \, \lltrans{\sigma_{i},P_i/\lambda_i} \, (q_{i+1}, \nu_{i+1}) \, \trans{\dots} \, (q_{j}, \nu_{j}) \lltrans{\sigma_{j},P_j/\lambda_j} \, (q_{j+1}, \nu_{j+1})
\]
then $C = (i, j, \mu)$ where $\dom(\mu) =\{X_1, \dots, X_k\} \subseteq L$. By definition, the automaton does the same as $\cA_\psi$, but it only considers the renamings in the transitions that marks the variables in $X_1, \dots, X_k \in L$. Finally, it satisfies the definition of the formula and $C \in \sem{\kPROJ_L(\psi)}(\Stream)$.

\noindent If $C \in \sem{\kPROJ_L(\psi)} (\Stream)$, then by definition there exists $C' \in \sem{\psi}(\Stream)$ such that $C = \kPROJ_L(C')$. This means that $C$ contains only the events that are in variables $X_i \in L$. As $C' \in \sem{\psi}(\Stream)$, then $C' \in \sem{\cA_{\psi}}(\Stream)$. On the other hand, a run of $\cA_\varphi$ does the same as a run of $\cA_\psi$ but the events that are marked in variables $X_i \in L$ are the only ones that are considered. Finally, it satisfies the definition of the automaton and $C \in \sem{\cA_\varphi}(\Stream)$.

\paragraph{[$\varphi = \kPROJ_{X(\ba_1, \dots, \ba_k)}(\psi)$]} If $\varphi = \kPROJ_{X(\ba_1, \dots, \ba_k)}(\psi)$, then $\cA_\varphi = (Q_\psi, \Delta_\varphi,q_{0\psi},F_\psi)$ where $\Delta_\varphi$ is the result of considering the renamings that are associated with the attributes $\ba_1, \dots, \ba_k$ in variable $X$ of each transition in $\Delta_\psi$. Formally, that is 
\begin{alignat*}{3}
	\qquad \Delta_\varphi \ = \ & \big\{(p,\sigma,P, \lambda' ,q) \mid && (p,\sigma,P,\lambda,q) \in \Delta_\psi &&\wedge \forall Y \neq X. \lambda'(Y) = \lambda(Y) \\
	& && \wedge \lambda'(X) = \bleft \ r' \mid && r \in \lambda(X) \\
	& && && \wedge \ \dom(r') = \dom(r) \cap \{\ba_1, \dots, \ba_k\} \\
	& && && \wedge \forall \ba_i \in \dom(r'). \ r'(\ba_i) = r(\ba_i) \bright\big\}.
\end{alignat*}
\noindent For a stream $\Stream = e_1 \ldots e_n$, assume that $\sem{\cA_{\psi}}(\Stream) = \sem{\psi}(\Stream)$. Then, we will prove that $\sem{\cA_\varphi}(\Stream) = \sem{\kPROJ_{X(\ba_1, \dots, \ba_k)}(\psi)}(\Stream)$.

\noindent Let $C \in \sem{\cA_\varphi}(\Stream)$, then an accepting run over $\Stream$ is of the form:
\[
\rho := (q_i, \nu_i) \, \lltrans{\sigma_{i},P_i/\lambda_i} \, (q_{i+1}, \nu_{i+1}) \, \trans{\dots} \, (q_{j}, \nu_{j}) \lltrans{\sigma_{j},P_j/\lambda_j} \, (q_{j+1}, \nu_{j+1})
\]
where $C_\rho = C = (i, j, \mu)$.
\noindent By definition, the automaton does the same as $\cA_\psi$, but it only considers the attributes $\{\ba_1, \dots, \ba_k\}$ of the renamings in the transitions that marks the variable $X$ and that have that attribute, and maintain the renamings of the other variables. Finally, it satisfies the definition of the formula and $C \in \sem{\kPROJ_{X(\ba_1, \dots, \ba_k)}(\psi)}(\Stream)$.

\noindent If $C \in \sem{\kPROJ_{X(\ba_1, \dots, \ba_k)}(\psi)} (\Stream)$ and $\Stream = e_1 \ldots e_n$, then by definition there exists $C' \in \sem{\psi}(\Stream)$ such that $C$ only considers the attributes $\{\ba_1, \dots, \ba_k\}$ in the events that are in variable $X$ and leave the other variables as they were from $C'$. As $C' \in \sem{\psi}(\Stream)$, then $C' \in \sem{\cA_{\psi}}(\Stream)$. On the other hand, a run of $\cA_\varphi$ does the same as a run of $\cA_\psi$ but only the attributes $\{\ba_1, \dots, \ba_k\}$ of the events that are marked in variable $X$ are considered, maintaining the other variables as they were. Finally, it satisfies the definition of the automaton and $C \in \sem{\cA_\varphi}(\Stream)$.

\paragraph{[$\varphi = \agg_{Y[\bb \gets \otimes X(\ba)]}(\psi)$]} Let $(\dset,\otimes, \zero)$ be the monoid associated with the aggregation.
Without loss of generality, we assume that the initial state $q_{0\psi}$ of $\cA_\psi$ has only \emph{outgoing transitions} and final states in $F_{\psi}$ have only \emph{incoming transitions}, namely, for every transition $(p,\sigma,P,\lambda,q) \in \Delta_\psi$ it holds that $q \neq q_{0\psi}$ and $p \notin F_{\psi}$. If not, we can add new states and use the non-determinism of \ACEA to satisfy these requirements. Further, for the sake of simplification, we will assume that from the initial state one cannot reach a final state \emph{in one step}, namely, for every transition $(q_{0\psi},\sigma,P,\lambda,q) \in \Delta_\psi$ it holds that $q \notin F_{\psi}$.
If this happens, one can easily extend the construction below to cover this case. 

So, assume that $\varphi = \agg_{Y[\bb \gets \otimes X(\ba)]}(\psi)$. We define $\cA_\varphi = (Q_\psi, \Delta_\varphi,q_{0\psi},F_\psi)$ where the new $\Delta_\varphi$ is the result extending $\Delta_{\psi}$ by adding a fresh register $\bc$ that is not being used anywhere in $\Delta_{\psi}$ and it will be used for the aggregation. Specifically, we define:
\[
\Delta_{\varphi} \ = \ \Delta_{\varphi}^{\operatorname{init}} \uplus \Delta_{\varphi}^{\operatorname{agg}} \uplus \Delta_{\varphi}^{\operatorname{out}}
\] 
where each set of transitions are defined as follows. 
First, the set $\Delta_{\varphi}^{\operatorname{init}}$ is in charged of initializing the register $c$ for performing the aggregation.
\begin{alignat*}{2}
	\quad \qquad \Delta_{\varphi}^{\operatorname{init}} = & \{(q_{0\psi},\sigma',P,\lambda,q) \mid && (q_{0\psi},\sigma,P,\lambda,q) \in \Delta_\psi \wedge \lambda(X) = \emptyset \wedge \sigma'(\bc) = \zero \\
	& && \wedge \forall \bd \neq \bc .\ \sigma'(\bd) = \sigma(\bd) \}\\
	& \cup \{(q_{0\psi},\sigma',P,\lambda,q) \mid && (q_{0\psi},\sigma,P,\lambda,q) \in \Delta_\psi \wedge \lambda(X) \neq \emptyset\\
	& && \wedge \sigma'(\bc) =
	\begin{cases}
		\nullvalue \ \ \ \ \text{ if } \exists r \in \lambda(X). \ r(\ba) = \nullvalue \\
		\bc \ \  \text{ otherwise} 
	\end{cases}  \\
	& && \wedge \forall \bd \neq \bc .\ \sigma'(\bd) = \sigma(\bd) \}
\end{alignat*}
Second, the set $\Delta_{\varphi}^{\operatorname{agg}}$ of transitions is in charged of the aggregation during the run and before reaching a final state.
\begin{alignat*}{2}
	\quad \qquad \Delta_{\varphi}^{\operatorname{agg}} = & \{(p,\sigma',P,\lambda,q) \mid && (p,\sigma,P,\lambda,q) \in \Delta_\psi \wedge \lambda(X) = \emptyset \wedge q \notin F \wedge \sigma'(\bc) = \bc \\
	& && \wedge \forall \bd \neq \bc .\ \sigma'(\bd) = \sigma(\bd) \}\\
	& \cup \{(p,\sigma',P,\lambda,q) \mid && (p,\sigma,P,\lambda,q) \in \Delta_\psi \wedge \lambda(X) \neq \emptyset \wedge q \notin F \\
	& && \wedge \sigma'(\bc) =
	\begin{cases}
		\nullvalue \ \ \ \ \text{ if } \exists r \in \lambda(X). \ r(\ba) = \nullvalue \\
		\bc \otimes \bigotimes_{\forall r \in \lambda(X)} \sigma(r(\ba)) \ \  \text{ otherwise} 
	\end{cases}  \\
	& && \wedge \forall \bd \neq \bc .\ \sigma'(\bd) = \sigma(\bd) \} 
\end{alignat*}
Finally, the set $\Delta_{\varphi}^{\operatorname{out}}$ has all transitions reaching a final state to produce the desire output of the aggregation.
\begin{alignat*}{2}
	\quad \qquad \Delta_{\varphi}^{\operatorname{out}} = & \ \{(p,\sigma',P,\lambda',q) \mid && (p,\sigma,P,\lambda,q) \in \Delta_\psi \wedge \lambda(X) = \emptyset \wedge q \in F \wedge \sigma'(\bc) = \bc \\
    & && \wedge \forall \bd \neq \bc .\ \sigma'(\bd) = \sigma(\bd) \\
    & && \wedge \lambda'(Y) = \lambda(Y) \cup \{[\bb \mapsto \bc]\} \\
	& && \wedge \forall Z \neq Y. \ \lambda'(Z) = \lambda(Z) \} \\
    & \cup \{(p,\sigma',P,\lambda',q) \mid && (p,\sigma,P,\lambda,q) \in \Delta_\psi \wedge \lambda(X) \neq \emptyset \wedge q \in F \\
	& && \wedge \sigma'(\bc) =
	\begin{cases}
		\nullvalue \ \ \ \ \text{ if } \exists r \in \lambda(X). \ r(\ba) = \nullvalue \\
		\bc \otimes \bigotimes_{\forall r \in \lambda(X)} \sigma(r(\ba)) \ \  \text{ otherwise} 
	\end{cases}  \\
	& && \wedge \forall \bd \neq \bc .\ \sigma'(\bd) = \sigma(\bd) \\
	& && \wedge \lambda'(Y) = \lambda(Y) \cup \{[\bb \mapsto \bc]\} \\
	& && \wedge \forall Z \neq Y. \ \lambda'(Z) = \lambda(Z)\}
\end{alignat*}
\noindent We assume that $\sem{\cA_{\psi}}(\Stream) = \sem{\psi}(\Stream)$. Next, we will prove that $$\sem{\cA_\varphi}(\Stream) = \sem{\agg_{Y[\bb \gets \otimes X(\ba)]}(\psi)}(\Stream).$$

\noindent Let $C \in \sem{\cA_\varphi}(\Stream)$, then an accepting run over a stream $\Stream = e_1 \ldots e_n$ is 
\[
\rho := (q_i, \nu_i) \, \lltrans{\sigma_{i},P_i/\lambda_i} \, (q_{i+1}, \nu_{i+1}) \, \trans{\dots} \, (q_{j}, \nu_{j}) \lltrans{\sigma_{j},P_j/\lambda_j} \, (q_{j+1}, \nu_{j+1})
\]
with $C = C_\rho = (i, j, \mu)$.
By definition, the automaton $\cA_\varphi$ does the same as $\cA_\psi$, but there is a new register $\bc$ in all transitions. In the first transition it sets $\bc$ as $\zero$ if $\lambda_0(X) \neq \emptyset$ or as $\bc$. Then, for each transition where $\lambda_i(X) \neq \emptyset$, it changes $\sigma_i$ to update $\bc$ with the value from the event and renamings, and for the others it maintains $\bc$. It also adds a new renaming in the transitions that go to a final state that puts the value of attribute $\bc$ in attribute $\bb$.
In summary, it satisfies the definition of the formula and $C \in \sem{\agg_{Y[\bb \gets \otimes X(\ba)]}(\psi)}$.

\noindent For the other direction, if $C \in \sem{\agg_{Y[\bb \gets \otimes X(\ba)]}(\psi)} (\Stream)$ and $\Stream = e_1 \ldots e_n$, then by definition of the formula, $C = (i, j,  \mu)$ where $\dom(\mu) = \{X_1, \dots, X_k, Y\}$ for some variables $X_1, \ldots, X_k \in \xset$ and there exists $C' \in \sem{\psi}(\Stream)$ over the same stream such that $C' = (i,j,\mu')$, $\mu'(X_i) = \mu(X_i)$, and $\mu(Y) = \mu'(Y) \cup \{e\}$ where $e$ is the new events with the aggregation. 

\noindent We also know that if $C' \in \sem{\psi}(\Stream)$ then $C' \in \sem{\cA_{\psi}}(\Stream)$. Then, we can find a run $\rho'$ of the automaton $\cA_{\psi}$ over $\Stream$ such that $C_{\rho'} = C'$. By the construction above, we can extend $\rho'$ to a run $\rho$ of $\cA_{\varphi}$ where the initial transition is taken from $\Delta_{\varphi}^{\operatorname{init}}$, the middle transitions from $\Delta_{\varphi}^{\operatorname{agg}}$, and the last transitions from $\Delta_{\varphi}^{\operatorname{out}}$. One can check that $\rho$ will additionally produce the event $e$ and $C = C_{\rho}$.

\paragraph{[$\varphi = \agg_{Y[\bb_1 \leftarrow \otimes_1 X_1(\ba_1), \ldots, \bb_\ell \leftarrow \otimes_\ell X_\ell(\ba_\ell)]}(\psi)$]} If $\varphi = \agg_{Y[\bb_1 \leftarrow \otimes_1 X_1(\ba_1), \ldots, \bb_\ell \leftarrow \otimes_\ell X_\ell(\ba_\ell)]}(\psi)$, then we can construct $\cA_\varphi$ analogously to the single-attribute aggregation, but adding $l$ new registers. The proof is also analogous.

\end{proof} 	
	\section{Examples from practice}\label{sec:all-examples}
In the following, we present several queries obtained from the literature and show how to model them with \acel. Given the operators introduced in the previous sections, recall that we define \acel as any formula $\varphi$ that uses the standard operators of \cel (Section~\ref{sec:preliminaries}), the aggregation operator $\agg$ (Section~\ref{sec:aggregation}), or a combination of them. We start this section by introducing some new operators and predicates that work as syntax sugar for \acel to define practical queries. Then we present three queries with aggregation obtained from three different CER proposals and specify them by using \acel.
\subsection{Useful operators in \acel for specifying real-life queries}

To specify CER queries in practice, the following operator will be useful. Let $R$ be any event type. We define the \acel formula $\nxt(R)$ with \cel such that:
\[
\nxt(R) \ \equiv \ \kPROJ_{R}\big( (X+:R) \kFILTER X[\type \neq R]\big) \kOR R
\]
where we assume that $R$ can also be used as a variable name (i.e., $R \in \xset$).
In other words, $\nxt(R)$ finds the first event of $R$-type in an interval and discard all other events in between.
As an example, one can use this operator to succinctly define the \cel query $S:\nxt(R)$ that finds all $S$-events directly followed by an $R$ event. 
We define this operator for its later use in examples appearing from other systems. 

Let $\ba$ be an attribute.
We also define some auxiliary predicates that will be use with the $\kFILTER$ operator to correlate two or more events. Recall that we define a predicate $P$ as a possibly infinite subset of events and we generalize $P$ from events to a multiset of events $E \subseteq \eset$ such that $E \models P$ if, and only if, $e \models P$ for every $e \in E$. For specifying queries in practice, we need some special multisets predicates that cannot be defined directly as a generalization of normal predicates. 
Formally, a \emph{multiset predicate} $P$ is a subset of multisets of events, namely, $P \subseteq \bagpow{\eset}$. For example, the generalization of an (event) predicate to multisets of events is a multiset predicate. 
These multisets predicates defined below will allow us to (1) ensure that the pattern has the same value at attribute $\ba$, (2) the value of an attribute is increasing, or (3) the value of an attribute is decreasing, respectively. More specifically, we define the following three multiset predicates. For a multiset of events $E$ and two events $e_1, e_2 \in E$, let $\mathsf{Succ}_E(e_1, e_2)$ be the logical formula that checks if $e_2$ is the \emph{successor} of $e_1$ in $E$, formally,  $e_1(\ctime) < e_2(\ctime)$ and there does not exist $e_3 \in E$ such that $e_1(\ctime) < e_3(\ctime)$ and $e_3(\ctime) < e_2(\ctime)$.
\begin{enumerate}
	\item We define the auxiliary predicate $[\ba]$ as:
	\[
	[\ba] \ := \ \{E \in \bagpow{\eset} \mid \forall e_1, e_2 \in E. \ e_1(\operatorname{\ba}) = e_2(\operatorname{\ba})\}
	\]
	\item We define the auxiliary predicate $[\operatorname{increasing}(\ba)]$ as:
	\[
	[\operatorname{increasing}(\ba)] \ := \{E \in \bagpow{\eset} \mid \forall e_1, e_2 \in E. \ \ \mathsf{Succ}_E(e_1, e_2) \rightarrow e_1(\operatorname{\ba}) < e_2(\operatorname{\ba})\}
	\]
	\item Finally, we define the auxiliary predicate $[\operatorname{decreasing}(\operatorname{\ba})]$ as:
	\[
	[\operatorname{decreasing}(\operatorname{\ba})] \ := \ \{E \in \bagpow{\eset} \mid \forall e_1, e_2 \in E. \ \ \mathsf{Succ}_E(e_1, e_2) \rightarrow e_1(\operatorname{\ba}) > e_2(\operatorname{\ba})\}
	\]
\end{enumerate}
Following \acel syntax and semantics, we use the above predicates with variables in~$\xset$. For example, we write $\varphi \kFILTER X[\ba]$ to define that all events in variable $X$ must satisfy~$[\ba]$. Similar as for standard predicates, we use conjunction and disjunction in $\kFILTER$ as a syntax sugar for composing filters or using $\kOR$, respectively.

In the following we provide several examples from previous literature and how we can specify them by using \acel. 

\begin{example}
	We use as an example an adaptation of `Query 1' query extracted from SASE's paper 
    \cite[p.2]{DiaoIG07}
    , which says: 
	\begin{quote}
		``Query 1 retrieves the total trading volume of Google stocks in the 4 hour period after some bad news occurred. The PATTERN clause declares the structure of a pattern. It uses the SEQ construct to specify a sequence pattern of two components: the first refers to an event whose type is news, and the second refers to a series of events of the stock type. The latter uses the Kleene plus, denoted by “+”, to represent one or more events of a particular type. A variable is declared in each component to refer to the corresponding event(s). A component that uses the Kleene plus declares its variable as an array using the “[ ]” symbols. The WHERE clause, if present, contains value-based radicates to define the events relevant to the pattern. In Query 1, the first predicate requires the type attribute of the news event to be bad. The second predicate requires every relevant stock event to have the symbol GOOG; the “every” semantics is expressed by b[i] (where i $\leq$ 1). We refer to such predicates as individual iterator predicates. The WITHIN clause specifies a window over the entire pattern, restricting the events considered to those within a 4 hour period. PATTERN, WHERE, and WITHIN clauses together completely define a pattern. Their evaluation over an event stream results in a stream of pattern matches. Each pattern match consists of a unique sequence of events used to match the pattern, stored in the a and b[ ] variables. The RETURN clause transforms each pattern match into a result event. In its specification, b[ ] implies an iterator over the events in the array, the volume attribute is retrieved from each event returned by the iterator, and the aggregate function sum() is applied to all the retrieved values.''
    \end{quote}
	The difference between this example and the original one is that this does not considers the within operator. The query in SASE's query language is the following:
	\begin{verbatim}
Q13: PATTERN SEQ(NEWS a, STOCK+ b[])
     WHERE a.type = `bad' and b[i].symbol = `GOOG'
     RETURN sum(b[].volume)
	\end{verbatim}
	A formula equivalent to the previous query \texttt{Q1} in \acel could be the following:
	\[
	\begin{array}{rcl}
		\varphi_{13} & = &  \agg_{Y[e \gets \sumtext (b(\operatorname{volume}))]}\big( \\
		& & \qquad [\mathsf{NEWS} \kAS a \fsq (\nxt (\mathsf{STOCK} \kAS b))\kSITER] \\
		& & \qquad \qquad \kFILTER (a[\operatorname{type} = \text{`bad'}] \wedge b[\operatorname{symbol} = \text{`GOOG'}])\big)
	\end{array}
	\]	
	Notice that the NEXT operator is necessary, since in SASE semantics the sequences are not contiguous, and we need to take exactly the one after an event NEWS. For this purpose, SASE uses what they call the NEXT selection strategy. Instead of modifying the semantics of our logic, we prefer to introduce a new operator that specifies this directly. 
\end{example}

\begin{example}
    We use as an example an adaptation of `Query 1' extracted from SASE's paper \cite[p.3]{SASEcomplexity}, which says
    \begin{quote}
        ``Query 1 computes the statistics of running times of mappers in Hadoop.
        The ‘Pattern’ clause specifies a seq pattern with three components: a single event indicating the start of a Hadoop job, followed by a Kleene+ for collating a series of events representing the mappers in the job, followed by an event marking the end of the job. Each component declares a variable to refer to the corresponding event(s), e.g, a, b[ ] and c, with the array variable b[ ] declared for Kleene+. The ‘Where’ clause uses these variables to specify value-based predicates. Here the predicates require all events to refer to the same job id; such equality comparison across all events can be writing with a shorthand, ‘[job id]’. The ‘Within’ clause specifies a 1-day window over the pattern. Finally, the Return’ clause constructs each output event to include the average and maximum durations of mappers in each job.`''
    \end{quote}
    The difference between this example and the original one is that this does not consider the within operators.
    \begin{Verbatim}[commandchars=\\\{\}]
Q14: PATTERN seq(JobStart a, Mapper+ b[ ], JobEnd c)
     WHERE a.job_id = b[i].job_id and a.job_id=c.job_id
     RETURN avg(b[ ].period), max(b[ ].period)
    \end{Verbatim}
    A formula equivalent to the previous query in CEL with aggregation could be the following:
    \[
        \begin{array}{rcl}
            \varphi_{14} & = & \agg_{Y[e \gets \max (b(\operatorname{period})), f \gets \avg (b(\operatorname{period}))]} \big(\\
            & & \qquad [(\mathsf{JobStart} \kAS a \kSEQ (\mathsf{Mapper} \kAS b) \kITER \kSEQ \mathsf{JobEnd} \kAS c) \kAS X]  \\
            & & \qquad \qquad \kFILTER \ X[\operatorname{job\_id}] \big) 
        \end{array}
    \]
\end{example}

\begin{example}\label{ex:esper}
    We use as an example an adaptation of a query extracted from ESPER's documentation \cite{espertech}, which says:
	\begin{quote}
		``This example statement demonstrates the idea by selecting a total price per customer over pairs of events (ServiceOrder followed by a ProductOrder event for the same customer id within 1 minute), occurring in the last 2 hours, in which the sum of price is greater than 100, and using a where clause to filter on name.''
	\end{quote}
	The difference between this example and the original one is that we do not consider group-by, window and slide operators and a slightly different filter. The query in ESPER's query language is the following:
\begin{Verbatim}[commandchars=\\\{\}]
Q15: SELECT a.custId, sum(a.price + b.price)
     FROM PATTERN [every a=ServiceOrder -> 
                           b=ProductOrder(custId = a.custId)]
     WHERE a.name in (b.name)
     HAVING sum(a.price + b.price) > 100
	\end{Verbatim}
    We can define \texttt{Q2} by using \acel as follows:
    \[
    \begin{array}{rcl}
    	\varphi_{15} & = & \Big[ \agg_{Y(e \gets \sumtext (X(price)))} \big(\\
    	& & \qquad [(\mathsf{ServiceOrder} \kAS a \kSEQ \mathsf{ProductOrder} \kAS b)\kITER \kAS X] \\
    	& & \qquad \qquad \kFILTER (X[\operatorname{name}]) \wedge \ X[\operatorname{custId}]\big)\Big]  \\
    	& & \qquad \qquad \qquad  \kFILTER Y[\operatorname{e} > 100]
    \end{array}
    \]
\end{example}

\begin{example}
    We use as an example an adaptation of a query extracted from GRETA's paper \cite[p.1]{Poppe2020}, which says
    \begin{quote}
        ``Query $Q_1$ computes the number of down-trends per sector during a time window of 10 minutes that slides every 10 seconds. These stock trends are expressed by the Kleene plus operator S+. All events in a trend carry the same company and sector identifier as required by the predicate [company, sector]. The predicate S.price \textgreater\ NEXT(S).price expresses that the price continually decreases from one event to the next in a trend. The query ignores local price fluctuations by skipping over increasing price records.''
    \end{quote}
    The difference between this example and the original one is that this does not considers slide, within and group-by operators.
    \begin{Verbatim}[commandchars=\\\{\}]
Q16: RETURN sector, COUNT(*) PATTERN Stock S+
     WHERE [company, sector] AND S.price > NEXT(S).price
    \end{Verbatim}
    A formula equivalent to the previous query in CEL with aggregation could be the following:
    \[
        \begin{array}{rcl}
        \varphi_{16} & = &  \agg_{Y [e \gets \cnt(S(\operatorname{id}))]} \big( [((\mathsf{Stock} \kAS S \kSEQ \mathsf{Stock}\kITER) \kAS X) \kITER]\\
        & & \qquad \kFILTER X[\operatorname{sector}] \wedge X[\operatorname{company}] \wedge X[\operatorname{decreasing}(\operatorname{sector})] \big)
        \end{array}
    \]
\end{example}

\begin{example}
    We use as an example an adaptation of query `$Q_2$' extracted from GRETA's paper \cite[p.1]{Poppe2020}, which says
    \begin{quote}
        ``Query $Q_2$ computes the total CPU cycles per job of each mapper experiencing increasing load trends on a cluster during a time window of 1 minute that slides every 30 seconds. A trend matched by the pattern of $Q_2$ is a sequence of a job-start event S, any number of mapper performance measurements M+, and a job-end event E. All events in a trend must carry the same job and mapper identifiers expressed by the predicate [job, mapper]. The predicate M.load \textless\ NEXT(M).load requires the load measurements to increase from one event to the next in a load distribution trend. The query may ignore any event to detect all load trends of interest for accurate cluster monitoring.''
    \end{quote}
    The difference between this example and the original one is that this does not considers slide, within and group-by operators.
    \begin{Verbatim}[commandchars=\\\{\}]
Q17: RETURN mapper, SUM(M.cpu)
     PATTERN SEQ(Start S, Measurement M+, End E)
     WHERE [job, mapper] AND M.load < NEXT(M).load
    \end{Verbatim}
    A formula equivalent to the previous query in CEL with aggregation could be the following:
    \[
        \begin{array}{rcl}
            \varphi_{17} & = & 
            \agg_{Y[e \gets \sumtext(M(\operatorname{cpu}))]}\big(\\
            & & \qquad [(\mathsf{Start} \kAS S \kSEQ \mathsf{Measurement} \kAS M+ \kSEQ \mathsf{End} \kAS E) \kAS X] \\
            & & \qquad \qquad \kFILTER \ X[\operatorname{job}] \wedge X[\operatorname{mapper}] \wedge M[\operatorname{increasing}(\operatorname{load})] \big) 
        \end{array}			
    \]
\end{example}

\begin{example}
    We use as an example an adaptation of the query `$q_1$' extracted from SHARON's paper \cite[p.1]{Poppe2018}, which says
    \begin{quote}
        ``Queries $q_1$ - $q_7$ in Figure 1 compute the count of trips on a route as a measure of route popularity.	They consume a stream of vehicle-position reports. Each report carries a time stamp in seconds, a car identifier and its position. Here, event type corresponds to a vehicle position. For example, a vehicle on Main Street sends a position report of type MainSt. Each trip corresponds to a sequence of position reports from the same vehicle (as required by the predicate [vehicle]) during a 10-minutes long time window that slides every minute (...) For example, pattern $p_1$ = (OakSt, MainSt) appears in queries $q_1$ - $q_4$. Sharing the aggregation of common patterns among multiple similar queries is vital to speed up system responsiveness.''
    \end{quote}
    The difference between this example and the original one is that this does not considers within and group-by operator.
    \begin{Verbatim}[commandchars=\\\{\}]
Q18: RETURN COUNT (*)
     PATTERN OakSt, MainSt, StateSt
     WHERE [vehicle]
    \end{Verbatim}
    A formula equivalent to the previous query in CEL with aggregation could be the following:
    \[
        \begin{array}{rcl}
            \varphi_{18} & = & 
            \agg_{Z[f \gets \cnt(Y(\operatorname{vehicle}))]}\big( \\
            & & \qquad [((\mathsf{OakSt} \kSEQ \mathsf{MainSt} \kSEQ \mathsf{StateSt} \kAS Y) \kAS X) \kITER]\\
            & & \qquad \qquad \kFILTER \ X[\operatorname{vehicle}]\big)
        \end{array}	
    \]
\end{example}

\begin{example}
    We use as an example an adaptation of query `$q_3$' extracted from HAMLET's paper \cite[p.1]{Poppe2021}, which says
    \begin{quote}
        ``All events in a trip must have the same driver and rider identifiers as required by the predicate [driver, rider] (...). Query $q_3$ tracks riders who cancel their accepted requests while the drivers were stuck in slow-moving traffic. All three queries contain the expensive Kleene sub-pattern T+ that matches arbitrarily long event trends.''`
    \end{quote}
    The difference between this example and the original one is that this does not considers within, slide and group-by operators.
    \begin{Verbatim}[commandchars=\\\{\}]
Q19: RETURN T.district, COUNT(*), SUM(T.duration)
     PATTERN SEQ(Request R, Travel T+, Cancel C)
     WHERE [driver, rider]  
    \end{Verbatim}
    A formula equivalent to the previous query in CEL with aggregation could be the following:
    \[
        \begin{array}{rcl}
            \varphi_{19} & = & 
            \agg_{Y [e \gets \sumtext(T(\operatorname{duration})), f \gets \cnt(C(\operatorname{driver}))]}\\
            & & \qquad [((\mathsf{Request} \kAS R \kSEQ \mathsf{Travel} \kAS T \kITER \kSEQ \mathsf{Cancel} \kAS C) \kAS X) \kITER] \\
            & & \qquad \qquad \kFILTER \ X[\operatorname{driver}] \wedge X[\operatorname{rider}]
        \end{array}
    \]
\end{example}

\begin{example}
    We use as an example an adaptation of a query extracted from COGRA's paper \cite[p.2]{Poppe2019}, which says
    \begin{quote}
        ``Query	$q_2$ computes the number of Uber pool trips that an Uber driver can complete when some riders cancel their trips after contacting the driver during a time window of 10 minutes that slides every 30 seconds. Each trip starts with a single Accept event, any number of Call and Cancel events, followed by a single Finish event. Each event carries a session identifier associated with the driver. All events that constitute one trip must carry the same session identifier as required by the predicate [driver]. The skip-till-next-match semantics allows query $q_2$ to skip irrelevant events such as in-transit, dropoff, etc.''`
    \end{quote}
    The difference between this example and the original one is that this does not considers within, slide and group-by operators.
    \begin{Verbatim}[commandchars=\\\{\}]
Q20: RETURN driver, COUNT(*)
     PATTERN SEQ(Accept, (SEQ(Call, Cancel))+, Finish)
     SEMANTICS skip-till-next-match
     WHERE [driver]
    \end{Verbatim}
    A formula equivalent to the previous query in CEL with aggregation could be the following:
    \[
        \begin{array}{rcl}
            \varphi_{20} & = & 
            \agg_{Z[f \gets \cnt(F(\operatorname{driver}))]} \big( \\
            & & \qquad [((\mathsf{Accept} \kSEQ (\mathsf{Call} \kSEQ \mathsf{Cancel}) \kITER \kSEQ \mathsf{Finish} \kAS F) \kAS X) \kITER]\\
            & & \qquad \qquad \kFILTER \ \mathsf{X[\operatorname{driver}]} \big)
        \end{array}
    \]
\end{example}

\begin{example}\label{ex:cogra}
    We use as an example an adaptation of the query `$q_1$' extracted from COGRA's paper \cite[p.1]{Poppe2019}, which says:
	\begin{quote}
		``Query $q_1$ detects minimal and maximal heartbeat during passive physical activities (e.g., reading, watching TV). Query $q_1$ consumes a stream of heart rate measurements of intensive care patients. Each event carries a time stamp in seconds, a patient identifier, an activity identifier, and a heart rate. For each patient, $q_1$ detects contiguously increasing heart rate measurements during a time window of 10 minutes that slides every 30 seconds. No measurements may be skipped in between matched events per patient, as expressed by the contiguous semantics.''
	\end{quote}
	The difference between this example and the original one is that this version does not considers within, group-by, and slide operators. The query is:
\begin{Verbatim}[commandchars=\\\{\}]
Q21: RETURN patient, MIN(M.rate), MAX(M.rate)
     PATTERN Measurement M+
     SEMANTICS contiguous
     WHERE [patient] AND M.rate < NEXT(M).rate 
     AND M.activity = passive
\end{Verbatim}
    An \acel formula equivalent to the previous query could be the following:
    \[
    \begin{array}{rcl}
    	\varphi_{21} & = & 
    	\agg_{Y [e \gets \min(M(\operatorname{rate})), f \gets \max(M(\operatorname{rate}))]}\big[ (\mathsf{Measurement} \kAS M) \kSITER \\
    	& & \qquad \kFILTER M[\operatorname{patient}] \wedge M[\operatorname{increasing}(\operatorname{rate})] \wedge M.\operatorname{activity} = \text{`passive'} \big]
    \end{array}
    \]
    One can check that formula $\varphi_3$ specifies the same query as \texttt{Q3} with the difference is that $\varphi_3$ has a formal and denotational semantics. 
\end{example}

It is important to note that we also consider examples of other proposals (e.g., CAYUGA \cite{demers2007cayuga}) that use aggregation; however, their query languages are procedural, and they do not adapt to the concept of declarative aggregation that we use in this work. 	
\end{document}